\documentclass[preprint,12pt,number,sort]{elsarticle}
\usepackage[T1]{fontenc}
\usepackage{lmodern}
\usepackage{amsmath,amssymb,amsthm,bm}
\usepackage{booktabs}
\usepackage{dcolumn}
\usepackage[a4paper,margin=25mm]{geometry}
\usepackage{xcolor}
\usepackage{graphicx}
\usepackage{tikz}
\usetikzlibrary{patterns,arrows.meta,positioning,calc}
\usepackage{microtype}
\usepackage{placeins}
\usepackage{etoolbox}
\patchcmd{\pprintMaketitle}{\unvbox\keybox\par\vskip10pt\fi\hrule\vskip12pt}{\unvbox\keybox\par\vskip10pt\fi}{}{\errmessage{pprintMaketitle patch failed}}
\usepackage[hidelinks]{hyperref}
\journal{Journal of Computational Physics}
\newcommand{\dd}{\,\mathrm d}
\newcommand{\vb}[1]{\bm{#1}}
\newcommand{\W}{\vb W}
\newcommand{\ms}[1]{\bm{#1}}
\newcommand{\tr}{\mathrm T}
\newcommand{\R}{\mathbb R}
\newcommand{\I}{I}
\newcommand{\M}{\mathcal M}
\newcommand{\Ecal}{\mathcal E}
\newcommand{\Iq}{\mathcal I}
\newcommand{\Aop}{\mathcal A}
\newcommand{\Vop}{\mathcal V}
\newcommand{\Gop}{\mathcal G}
\newcommand{\Feq}{\vb F^{\rm eq}}
\newcommand{\Ffw}{\vb F^{{\rm fr},{\rm wave}}}
\newcommand{\Ffp}{\vb F^{{\rm fr},p}}
\newcommand{\Wfp}{\W^{{\rm fr},p}}
\newtheorem{proposition}{Proposition}
\newcolumntype{d}[1]{D{.}{.}{#1}}
\newcommand{\colhead}[1]{\multicolumn{1}{c}{#1}}
\AtBeginEnvironment{table}{\setlength{\belowcaptionskip}{6pt}}

\begin{document}
\begin{frontmatter}

\title{A semi-implicit unified gas-kinetic wave--particle method with energy-conserving electromagnetic coupling for multiscale plasmas}

\author[a]{Yixiao Wang}
\ead{ywangyp@connect.ust.hk}
\author[a]{Zhigang Pu}
\ead{zpuac@connect.ust.hk}
\author[c]{Xing Ji}
\ead{jixing@xjtu.edu.cn}
\author[a,b]{Kun Xu\corref{cor1}}
\ead{makxu@ust.hk}
\cortext[cor1]{Corresponding author.}

\affiliation[a]{organization={Department of Mathematics, Hong Kong University of Science and Technology},
            addressline={Clear Water Bay},
            city={Kowloon},
            country={Hong Kong}}
\affiliation[b]{organization={Shenzhen Research Institute, Hong Kong University of Science and Technology},
            city={Shenzhen},
            country={China}}
\affiliation[c]{organization={Shaanxi Key Laboratory of Environment and Control for Flight Vehicle, Xi'an Jiaotong University},
            city={Xi'an},
            country={China}}

\begin{abstract}
In multiscale plasmas, kinetic regions coexist with collision-dominated, strongly magnetized and quasi-neutral ones. In these
regions, the collision time, the cyclotron period and the plasma period, respectively, are far shorter than the flow time
scales. Schemes that treat the collisions and the field coupling explicitly must take time steps below these times and below
the light-crossing time of a cell. This paper presents a semi-implicit unified gas-kinetic wave--particle (UGKWP) method for the multispecies Vlasov--Maxwell
system with Bhatnagar--Gross--Krook collisions. The electric field follows from a reformulated Maxwell equation, a single linear system with a unique solution for any
time step. Its current response is built from the implicit-midpoint velocity map, which then pushes the particles and the
hydrodynamic wave alike. The field step conserves the discrete total energy exactly and remains stable without resolving the cyclotron period, the plasma period or the light-crossing time, so that the time step follows from the flow and sound speeds. A Gauss's law correction adjusts the electric field and the joint
charge of particles and wave while conserving the mass, momentum and energy of every charged species. The
method reproduces whistler waves and electron plasma waves from Landau damping to collision-dominated oscillations. It agrees with particle-in-cell, two-fluid and magnetohydrodynamic references in a
Brio--Wu problem from weak to strong collisionality and magnetization, the Orszag--Tang vortex and collisionless
magnetic reconnection. It needs far less CPU time than either the UGKWP method with an explicit solver for the perfectly hyperbolic Maxwell
equations or a deterministic multiscale kinetic scheme. In collisional plasmas, it also needs less CPU time than an implicit
particle-in-cell method with the same collision operator, increasingly so as collisions strengthen.
\end{abstract}


\end{frontmatter}
\section{Introduction}\label{sec:intro}

The collisionality and magnetization of a plasma can vary by orders of magnitude within a single domain. Where the
Knudsen number is large, the species require a kinetic description. Where it is small, they relax to local equilibrium
and are governed by multi-fluid equations, which reduce in turn to ideal magnetohydrodynamics (MHD) in the limit of
small normalized Larmor radius and Debye length \cite{Liu2021}. Discrete-velocity methods that split transport and
collisions must resolve the mean free path and the collision time to keep their numerical dissipation below the physical
one \cite{Xu2010,Dimarco2014}. Explicit particle-in-cell (PIC)
methods must resolve the Debye length, the plasma period and the cyclotron period, and an explicit field solver also
bounds the time step by the time in which light crosses a cell \cite{Dawson1983}. In the collision-dominated and MHD
regimes, these scales are far smaller than those of the flow.

For the collisional scales, asymptotic-preserving (AP) schemes capture the continuum limit without resolving the
collision time \cite{Jin1999,Dimarco2014}. The unified gas-kinetic scheme (UGKS) \cite{Xu2010} and the
discrete unified gas-kinetic scheme (DUGKS) \cite{Guo2013} construct their interface fluxes from the
integral solution of the kinetic equation, which couples free transport and collisions within a time step. They thus capture the flow physics from free transport to the Navier--Stokes solutions according to the local cell Knudsen
number, without resolving the mean free path and the collision time.
Both have been extended to multicomponent plasmas coupled to Maxwell's equations \cite{Liu2017,LiuDUGKS2026}. The unified gas-kinetic wave--particle (UGKWP) method
\cite{Liu2020,Zhu2019} retains the multiscale flux of UGKS without its discrete velocity grid. Stochastic particles
carry the part of the distribution function that streams freely over a time step, and an analytic hydrodynamic wave
carries the part that collides within it. The particle fraction therefore decreases as the collision
frequency increases and vanishes in the continuum limit.

For the electromagnetic scales, implicit PIC methods remove the stability limits that the plasma period and the
light-crossing time impose on the time step \cite{Brackbill1982,Markidis2010}. The energy-conserving semi-implicit
method (ECsim) \cite{Lapenta2017} conserves the total energy exactly with a single linear system for the fields per
time step, by building the current response from the same velocity update that then pushes the particles. Exact
conservation of both the energy and the charge has been achieved in implicit electrostatic \cite{Chen2011} and
semi-implicit electromagnetic finite-element \cite{CamposPinto2022} PIC methods. As the Debye length tends to zero, the plasma becomes quasi-neutral, and Maxwell's equations no longer determine the
electric field. The reformulated Maxwell equation (RME) remains well posed in this limit, and particle methods based on
it capture the quasi-neutral limit without resolving the plasma period and the Debye length \cite{Degond2017a}. AP particle methods have since been developed that conserve the discrete energy and satisfy Gauss's law, either by correcting the electric field \cite{Ji2023} or by conserving the charge \cite{Ma2024}. For grid-based
kinetic schemes, the RME has been used in a conservative semi-Lagrangian method that is AP in the quasi-neutral limit
\cite{Liu2025} and in DUGKS \cite{LiuDUGKS2026} for the Vlasov--Maxwell system with Bhatnagar--Gross--Krook (BGK)
collisions \cite{Bhatnagar1954}. The latter captures both the
continuum and the quasi-neutral limits. With a reformulated Poisson equation, UGKS and UGKWP capture the same limits
for electrostatic plasmas \cite{Pu2025b}.

The UGKWP method of Liu and Xu \cite{Liu2021} advances Maxwell's equations by an unconditionally stable Crank--Nicolson
scheme and couples the momentum equations implicitly to the electromagnetic field. In the highly magnetized regime, its cell size and time
step are then not restricted by the Debye length and the cyclotron period. The UGKWP method of Pu and
Xu \cite{Pu2025a} advances Maxwell's equations explicitly in their perfectly hyperbolic form \cite{Munz2000}, with a time
step limited by the speed of light, and couples the macroscopic variables of the species to the field by a Crank--Nicolson step. In UGKWP, the particles carry about the fraction $e^{-\Delta t/\tau}$ of the distribution function over a time step
$\Delta t$, where $\tau$ is the relaxation time. Even in collision-dominated plasmas, $\tau$ can far exceed the plasma
period and the light-crossing time. A time step bounded by these times then leaves nearly
the whole distribution function to the particles, and the wave--particle representation saves little over a particle
method. A field coupling free of these restrictions is therefore needed. Over longer time steps, the field and the species exchange energy through the current
in the field equation. As in ECsim, this exchange is balanced exactly if the current results from the same velocity
update that accelerates the charges. The UGKWP methods of \cite{Liu2021} and \cite{Pu2025a}, however, take the current
from the macroscopic momentum equations and accelerate the particles in a separate update. Since the analytic wave carries part of the current, the velocity update that balances the exchange
has to act on the particles and the wave alike.

Besides the energy balance, the discrete fields should satisfy Gauss's law. With this coupling, the current in the field equation results from the velocity update, whereas the charge moves only in the transport steps. This current does not satisfy the discrete continuity equation, so that Gauss's law does not hold by construction, unlike in
PIC methods with charge-conserving current deposition \cite{Esirkepov2001}
and in compatible particle--field discretizations \cite{Kraus2017}. Other methods control the Gauss defect by correcting
the electric field \cite{Marder1987,Munz2000,LiuDUGKS2026} or the particle positions \cite{Chen2019}. In quasi-neutral regions, a correction of the electric field at fixed charge
can turn a small charge error into a large change of the field. Since part of the charge belongs to the wave, which is
converted to and from particles in every transport step, a correction of the charge in UGKWP has to act on the particles and the wave together.

This paper presents UGKWP-EC, a semi-implicit UGKWP method for the multispecies BGK--Vlasov--Maxwell system, with an
electromagnetic coupling that conserves energy. Between two UGKWP
transport steps, a field step determines the electric field from an RME, a single linear system with a unique
solution for any time step. Its current response is built from the implicit-midpoint velocity map, which then pushes the
particles and the wave alike. Together with the implicit midpoint rule for Maxwell's equations, this common
map makes the field step conserve the discrete total energy exactly.
At the end of each time step, a Gauss's law correction adjusts the electric field and the masses of
the particles and the wave while conserving the mass, momentum and energy of every charged species.

The field step remains stable without resolving the cyclotron period, the plasma period or the light-crossing
time. The tests take the time
step from a CFL condition on the flow and sound speeds of the species and use PIC, two-fluid and MHD references. In one dimension, they cover plasma waves, one of them with a collision frequency that varies
over four orders of magnitude in space, and a Brio--Wu problem from kinetic plasmas to the MHD limit. In two dimensions, they cover
the Orszag--Tang vortex and collisionless magnetic reconnection. In both, UGKWP-EC, whose time step does not depend on the speed of light, needs less CPU time than the
UGKWP method of \cite{Pu2025a}. In the Orszag--Tang vortex, it also needs less CPU time than the implicit PIC code
iPIC3D \cite{Markidis2010} with the same collision operator, and the saving grows as the collisions strengthen.
In collisionless magnetic reconnection, it agrees with DUGKS \cite{LiuDUGKS2026} at less than one percent of the CPU time.

The remainder of the paper is organized as follows. Section~\ref{sec:model} states the BGK--Vlasov--Maxwell system, and
Section~\ref{sec:ugkwp} reviews the UGKWP method. Section~\ref{sec:method} presents UGKWP-EC and its discrete
properties, Section~\ref{sec:tests} reports the numerical tests and the computational cost, and Section~\ref{sec:conclusions} concludes the paper.

\section{Governing equations}\label{sec:model}

Consider $N_s$ monatomic species with masses $m_s$, charges $q_s$ and charge-to-mass ratios
$\kappa_s=q_s/m_s$. The evolution of the distribution function $f_s(\vb x,\vb u,t)$ of species $s$, with position
$\vb x\in\Omega\subset\R^d$ and microscopic velocity $\vb u\in\R^3$, is governed by the kinetic equation
\begin{equation}
 \partial_t f_s+\vb u\cdot\nabla_{\vb x} f_s+\kappa_s(\vb E+\vb u\times\vb B)\cdot\nabla_{\vb u} f_s
 =\frac{\M[\W_s]-f_s}{\tau_s},\qquad s=1,\dots,N_s,
 \label{eq:kinetic}
\end{equation}
where $\vb E$ and $\vb B$ are the electric and magnetic fields and the right-hand side is the BGK operator
\cite{Bhatnagar1954}, which relaxes $f_s$ with the relaxation time $\tau_s$ toward the Maxwellian $\M[\W_s]$ defined
below. The conservative variables of species $s$ are its mass, momentum and total energy densities,
\begin{equation*}
 \W_s=\int_{\R^3}\vb\psi\,f_s\dd\vb u=(\rho_s,\vb P_s,\Ecal_s)^\tr,\qquad
 \vb\psi(\vb u)=\bigl(1,\vb u,\tfrac12|\vb u|^2\bigr)^\tr.
\end{equation*}
Its number density, macroscopic velocity and temperature are
\begin{equation*}
 n_s=\frac{\rho_s}{m_s},\qquad \vb U_s=\frac{\vb P_s}{\rho_s},\qquad T_s=\frac{2}{3n_sk_B}\Bigl(\Ecal_s-\frac12\rho_s|\vb U_s|^2\Bigr),
\end{equation*}
where $k_B$ is the Boltzmann constant, and its Maxwellian distribution is
\begin{equation*}
 \M[\W_s](\vb u)=\rho_s\Bigl(\frac{m_s}{2\pi k_BT_s}\Bigr)^{3/2}\exp\Bigl(-\frac{m_s}{2k_BT_s}|\vb u-\vb U_s|^2\Bigr).
\end{equation*}

The relaxation time of species $s$ is $\tau_s=1/\nu_s$, with the collision frequency $\nu_s=\chi_{ss}n_s$. The
coefficient $\chi_{ss}\ge0$ of the collisions within species $s$ is a function of $\vb x$ and of the local state
$\W_s$.

The electromagnetic fields evolve according to Maxwell's equations
\begin{equation}
 \epsilon_0\partial_t\vb E=\mu_0^{-1}\nabla\times\vb B-\vb J,\qquad
 \partial_t\vb B=-\nabla\times\vb E,\qquad
 \vb J=\sum_s\kappa_s\vb P_s,
 \label{eq:maxwell}
\end{equation}
with the permittivity $\epsilon_0$, the permeability $\mu_0$ and the speed of light $c=(\epsilon_0\mu_0)^{-1/2}$. The
fields satisfy Gauss's law $\epsilon_0\nabla\cdot\vb E=\rho_c$, with the charge density $\rho_c=\sum_s\kappa_s\rho_s$,
and $\nabla\cdot\vb B=0$. With \eqref{eq:maxwell}, the energy moments of \eqref{eq:kinetic}, summed over the species,
give the conservation law of the total energy,
\begin{equation*}
 \partial_t\Bigl(\sum_s\Ecal_s+\frac{\epsilon_0}{2}|\vb E|^2+\frac{1}{2\mu_0}|\vb B|^2\Bigr)
 +\nabla\cdot\Bigl(\sum_s\int_{\R^3}\frac12|\vb u|^2\vb u f_s\dd\vb u+\frac{1}{\mu_0}\vb E\times\vb B\Bigr)=0 .
\end{equation*}

\label{sec:model-units}The equations are nondimensionalized by the reference length $x_0$, time $t_0$, mass density $\rho_0$,
mass $m_0$ and magnetic field $B_0$, with the reference velocity $u_0=x_0/t_0$ and electric field $E_0=u_0B_0$.
The nondimensional variables, denoted by a
prime, are defined by
\[
 \begin{gathered}
 \vb x=x_0\vb x',\quad t=t_0t',\quad \vb u=u_0\vb u',\quad f_s=\frac{\rho_0}{u_0^3}f_s',\quad
 (\rho_s,\vb P_s,\Ecal_s)=\bigl(\rho_0\rho_s',\rho_0u_0\vb P_s',\rho_0u_0^2\Ecal_s'\bigr),\\
 m_s=m_0m_s',\quad n_s=\frac{\rho_0}{m_0}n_s',\quad T_s=\frac{m_0u_0^2}{k_B}T_s',\quad \vb E=E_0\vb E',\quad \vb B=B_0\vb B'.
 \end{gathered}
\]
In these variables, \eqref{eq:kinetic}, \eqref{eq:maxwell}, Gauss's law and $\nabla\cdot\vb B=0$ keep their form,
with $k_B=1$ and the coefficients
\[
 \kappa_s'=\kappa_sB_0t_0,\qquad
 \epsilon_0'=\frac{\epsilon_0B_0^2}{\rho_0},\qquad
 \frac1{\mu_0'}=\frac{B_0^2}{\mu_0\rho_0u_0^2},\qquad
 \chi_{ss}'=\frac{\chi_{ss}\rho_0t_0}{m_0}.
\]

\section{The unified gas-kinetic wave--particle method}\label{sec:ugkwp}

This section briefly reviews the UGKWP method \cite{Liu2020,Zhu2019}
for a single species without electromagnetic fields, governed by the BGK equation
\begin{equation}
 \partial_tf+\vb u\cdot\nabla_{\vb x}f=\frac{g-f}{\tau},\qquad g=\M[\W],\qquad \W=\int_{\R^3}\vb\psi f\dd\vb u,
 \label{eq:bgk}
\end{equation}
with the relaxation time $\tau$.

\subsection{Finite-volume framework and interface flux}\label{sec:ugkwp-fv}

The domain is divided into cells $\Omega_i$ of volume $|\Omega_i|$. The interface $S_{ij}$
between $\Omega_i$ and an adjacent cell $\Omega_j$, $j\in N(i)$, has the area $|S_{ij}|$, the center $\vb x_{ij}$
and the unit normal $\vb n_{ij}$ pointing into $\Omega_j$. Over a time step $\Delta t=t^{n+1}-t^n$, the cell averages
$\W_i$ are updated in conservation form with the time-integrated fluxes per unit area
$\vb F_{ij}=\int_{t^n}^{t^{n+1}}\int_{\R^3}\vb u\cdot\vb n_{ij}\,\vb\psi\,f(\vb x_{ij},\vb u,t)\dd\vb u\dd t$
\cite{Zhu2019}. In UGKS \cite{Xu2010}, the distribution function at the interface is the
integral solution of \eqref{eq:bgk} along the characteristics,
\begin{equation}
 f(\vb x_{ij},\vb u,t)=\frac1\tau\int_{t^n}^te^{-(t-t')/\tau}g\bigl(\vb x_{ij}-\vb u(t-t'),\vb u,t'\bigr)\dd t'
 +e^{-(t-t^n)/\tau}f^n\bigl(\vb x_{ij}-\vb u(t-t^n),\vb u\bigr),
 \label{eq:ugkwp-integral}
\end{equation}
where $f^n$ is the distribution function at $t^n$. The equilibrium is expanded about $\vb x_{ij}$ and $t^n$ as $g=g_0+(\vb x-\vb x_{ij})\cdot\vb g_{\vb x}+(t-t^n)g_t$, and the initial
distribution function in the upwind cell, $\Omega_i$ for $\vb u\cdot\vb n_{ij}>0$ and $\Omega_j$ for $\vb u\cdot\vb n_{ij}<0$, as
$f^n=f_0+(\vb x-\vb x_{ij})\cdot\vb f_{\vb x}$. The flux is then the sum $\vb F_{ij}=\Feq_{ij}+\vb F^{\rm fr}_{ij}$ of the equilibrium flux and the
free-transport flux,
\begin{equation}
 \begin{aligned}
 \Feq_{ij}&=\int_{\R^3}\vb u\cdot\vb n_{ij}\,\vb\psi\,\bigl(q_1g_0+q_2\vb u\cdot\vb g_{\vb x}+q_3g_t\bigr)\dd\vb u,\\
 \vb F^{\rm fr}_{ij}&=\int_{\R^3}\vb u\cdot\vb n_{ij}\,\vb\psi\,\bigl(q_4f_0+q_5\vb u\cdot\vb f_{\vb x}\bigr)\dd\vb u,
 \end{aligned}
 \label{eq:ugkwp-flux}
\end{equation}
with the coefficients
\begin{equation*}
 \begin{alignedat}{2}
 q_1&=\Delta t-\tau\bigl(1-e^{-\Delta t/\tau}\bigr),&\qquad
 q_4&=\tau\bigl(1-e^{-\Delta t/\tau}\bigr),\\
 q_2&=2\tau^2\bigl(1-e^{-\Delta t/\tau}\bigr)-\tau\Delta t-\tau\Delta t\,e^{-\Delta t/\tau},&\qquad
 q_5&=\tau\Delta t\,e^{-\Delta t/\tau}-\tau^2\bigl(1-e^{-\Delta t/\tau}\bigr),\\
 q_3&=\frac{\Delta t^2}2-\tau\Delta t+\tau^2\bigl(1-e^{-\Delta t/\tau}\bigr).
 \end{alignedat}
\end{equation*}
The equilibrium at the interface is formed from the reconstructed conservative variables,
\[
 \begin{gathered}
 g_0=\M[\W_0],\qquad
 \W_0=\int_{\vb u\cdot\vb n_{ij}>0}\vb\psi\,g_L\dd\vb u+\int_{\vb u\cdot\vb n_{ij}<0}\vb\psi\,g_R\dd\vb u,\\
 \W_{\vb x}=\int_{\vb u\cdot\vb n_{ij}>0}\vb\psi\,\nabla g_L\dd\vb u+\int_{\vb u\cdot\vb n_{ij}<0}\vb\psi\,\nabla g_R\dd\vb u,
 \end{gathered}
\]
where $g_L$ and $g_R$ are the Maxwellians of the conservative variables reconstructed at $\vb x_{ij}$ in $\Omega_i$ and
$\Omega_j$, and $\W_{\vb x}$ is the spatial derivative of $\W_0$. The derivatives $\vb g_{\vb x}$ and $g_t$ follow from
$\W_{\vb x}$ and the compatibility condition,
\[
 \int_{\R^3}\vb\psi\,\vb g_{\vb x}\dd\vb u=\W_{\vb x},\qquad
 \int_{\R^3}\vb\psi\,g_t\dd\vb u=-\int_{\R^3}\vb\psi\,\vb u\cdot\vb g_{\vb x}\dd\vb u.
\]

\subsection{Wave--particle decomposition}\label{sec:ugkwp-wave}

The distribution function is represented in part by particles $P_k(m_k,\vb x_k,\vb u_k)$ with mass $m_k$, position $\vb x_k$
and velocity $\vb u_k$. By \eqref{eq:ugkwp-integral}, a particle streams freely for a time $t$ with probability $e^{-t/\tau}$ and has
collided otherwise. Accordingly, every particle is given the
free-transport time
\begin{equation}
 t_{f,k}=\min\bigl(-\tau\ln\eta_k,\ \Delta t\bigr),
 \label{eq:ugkwp-tf}
\end{equation}
with $\eta_k$ uniformly distributed on $(0,1)$, and moves freely to $\vb x_k+\vb u_kt_{f,k}$. The net moments that the
particles carry into $\Omega_i$ during their free transport are
\begin{equation}
 \Wfp_i=\sum_{\vb x_k+\vb u_kt_{f,k}\in\Omega_i}m_k\vb\psi(\vb u_k)-\sum_{\vb x_k\in\Omega_i}m_k\vb\psi(\vb u_k).
 \label{eq:ugkwp-pflux}
\end{equation}
Particles with $t_{f,k}=\Delta t$ are collisionless. Collisional particles, with $t_{f,k}<\Delta t$, are removed at the
end of the step. The part of the distribution function that the remaining, collisionless particles do not represent is the hydrodynamic
wave
\begin{equation*}
 \W^h_i=\W_i-\W^p_i,\qquad
 \W^p_i=\frac1{|\Omega_i|}\sum_{\vb x_k\in\Omega_i}m_k\vb\psi(\vb u_k).
\end{equation*}
Of the hydrodynamic wave, only the part
\begin{equation}
 \W^{hp}_i=e^{-\Delta t/\tau}\W^h_i
 \label{eq:ugkwp-whp}
\end{equation}
moves freely over the whole step. This part is sampled from the Maxwellian of the reconstructed hydrodynamic wave into particles whose free-transport time is $\Delta t$. The remaining part,
$(1-e^{-\Delta t/\tau})\W^h_i$, collides within the step. Its free transport before the collision is computed
analytically as the free-transport flux \eqref{eq:ugkwp-flux} of the wave without the
part that the sampled particles carry,
\begin{equation}
 \Ffw_{ij}=\int_{\R^3}\vb u\cdot\vb n_{ij}\,\vb\psi\,\Bigl[\bigl(q_4-\Delta t\,e^{-\Delta t/\tau}\bigr)g^h_0
 +\Bigl(q_5+\frac{\Delta t^2}{2}e^{-\Delta t/\tau}\Bigr)\vb u\cdot\vb g^h_{\vb x}\Bigr]\dd\vb u,
 \label{eq:ugkwp-frwave}
\end{equation}
where $g^h_0$ and $\vb g^h_{\vb x}$ are the Maxwellian of the reconstructed hydrodynamic wave at $\vb x_{ij}$ and its
spatial derivative.
The cell averages are updated by
\begin{equation}
 \W_i^{n+1}=\W_i^n-\frac1{|\Omega_i|}\sum_{j\in N(i)}\bigl(\Feq_{ij}+\Ffw_{ij}\bigr)|S_{ij}|+\frac{\Wfp_i}{|\Omega_i|},
 \label{eq:ugkwp-update}
\end{equation}
where $\Wfp_i$ includes the sampled particles.

\label{sec:ugkwp-algorithm}One time step consists of the following steps (Fig.~\ref{fig:ugkwp}).
\begin{description}
\item[Step 1.] Form the hydrodynamic wave $\W^h_i=\W^n_i-\W^p_i$ from the cell averages and the particles that remain
from the previous step (Fig.~\ref{fig:ugkwp}(a)).
\item[Step 2.] Obtain the free-transport time \eqref{eq:ugkwp-tf} of every remaining particle
(Fig.~\ref{fig:ugkwp}(b)). Sample the collisionless particles of each cell from its hydrodynamic wave, with the expected moments $|\Omega_i|\W^{hp}_i$
\eqref{eq:ugkwp-whp} and the free-transport time $\Delta t$ (Fig.~\ref{fig:ugkwp}(c)).
\item[Step 3.] Stream all particles freely for their free-transport times, accumulate the net particle moments
$\Wfp_i$ \eqref{eq:ugkwp-pflux}, and compute the fluxes $\Feq_{ij}$ \eqref{eq:ugkwp-flux} and $\Ffw_{ij}$
\eqref{eq:ugkwp-frwave}.
\item[Step 4.] Update the cell averages $\W^{n+1}_i$ by \eqref{eq:ugkwp-update} and remove the collisional particles
(Fig.~\ref{fig:ugkwp}(d)).
\end{description}

\begin{figure}[!tbp]
\centering
\begin{tikzpicture}[font=\footnotesize,
  wave/.style={pattern=north east lines,pattern color=black!50},
  pf/.style={draw,circle,inner sep=0pt,minimum size=4.6pt,fill=white},
  pc/.style={draw,circle,inner sep=0pt,minimum size=4.6pt,fill=black},
  lab/.style={fill=white,inner sep=1.5pt}]
 \def\L{3.0}\def\H{2.6}\def\xs{2.0}\def\ys{0.75}
 \def\xl{0.25,0.6,0.95,1.3,1.65}\def\xr{2.2,2.5,2.8}
 \begin{scope}
  \fill[wave] (0,\ys) rectangle (\L,\H);
  \foreach \x in \xl \foreach \y in {0.22,0.53} \node[pf] at (\x,\y) {};
  \foreach \x in \xr \foreach \y in {0.22,0.53} \node[pf] at (\x,\y) {};
  \draw (0,0) rectangle (\L,\H); \draw (0,\ys) -- (\L,\ys);
  \node[lab] at (1.5,1.67) {$\W^h$};
  \node[anchor=east] at (-0.03,0.375) {$\W^p$};
  \node at (1.5,-0.55) {(a)};
 \end{scope}
 \begin{scope}[xshift=3.7cm]
  \fill[wave] (0,\ys) rectangle (\L,\H);
  \foreach \x in \xl \foreach \y in {0.22,0.53} \node[pc] at (\x,\y) {};
  \foreach \x in \xr \foreach \y in {0.22,0.53} \node[pf] at (\x,\y) {};
  \draw (0,0) rectangle (\L,\H); \draw (0,\ys) -- (\L,\ys);
  \draw[densely dashed] (\xs,0) -- (\xs,\H);
  \node[lab] at (1.5,1.67) {$\W^h$};
  \node at (1.0,\H+0.3) {$1-e^{-\Delta t/\tau}$};
  \node at (2.5,\H+0.3) {$e^{-\Delta t/\tau}$};
  \node at (1.0,-0.2) {\scriptsize $t_f<\Delta t$};
  \node at (2.5,-0.2) {\scriptsize $t_f=\Delta t$};
  \node at (1.5,-0.55) {(b)};
 \end{scope}
 \begin{scope}[xshift=7.4cm]
  \fill[wave] (0,\ys) rectangle (\xs,\H);
  \foreach \x in \xl \foreach \y in {0.22,0.53} \node[pc] at (\x,\y) {};
  \foreach \x in \xr \foreach \y in {0.22,0.53,1.00,1.30,1.60} \node[pf] at (\x,\y) {};
  \draw (0,0) rectangle (\L,\H); \draw (0,\ys) -- (\L,\ys);
  \draw[densely dashed] (\xs,0) -- (\xs,\H);
  \node[lab] at (1.0,1.67) {$\W^h-\W^{hp}$};
  \node at (2.5,2.12) {$\W^{hp}$};
  \node at (1.0,\H+0.3) {$1-e^{-\Delta t/\tau}$};
  \node at (2.5,\H+0.3) {$e^{-\Delta t/\tau}$};
  \node at (1.5,-0.55) {(c)};
 \end{scope}
 \begin{scope}[xshift=11.1cm]
  \fill[wave] (0,0) rectangle (\xs,\H);
  \foreach \x in \xr \foreach \y in {0.22,0.53,1.00,1.30,1.60} \node[pf] at (\x,\y) {};
  \draw (0,0) rectangle (\L,\H);
  \draw[densely dashed] (\xs,0) -- (\xs,\H);
  \node[lab] at (1.0,1.3) {$\W^h$};
  \node[anchor=west] at (\L+0.03,1.3) {$\W^p$};
  \node at (1.5,-0.55) {(d)};
 \end{scope}
\end{tikzpicture}
\caption{Composition of the distribution function in one step of the UGKWP method.
The hatched region is the hydrodynamic wave, hollow circles are collisionless particles and solid circles are
collisional particles. (a) Beginning of the step. (b) Classification of the particles by their free-transport times.
(c) Sampling of $\W^{hp}$ from the hydrodynamic wave. (d) End of the step, after the removal of the collisional
particles. The widths
$1-e^{-\Delta t/\tau}$ and $e^{-\Delta t/\tau}$ indicate the collisional and the collisionless parts.}
\label{fig:ugkwp}
\end{figure}
\section{Wave--particle method for the BGK--Vlasov--Maxwell system}\label{sec:method}

\subsection{Time splitting and discrete state}\label{sec:split}

The system \eqref{eq:kinetic}, \eqref{eq:maxwell} is split into a transport--relaxation operator
and a field operator,
\begin{equation*}
 \begin{aligned}
 \Aop:&\quad \partial_tf_s+\vb u\cdot\nabla_{\vb x}f_s=\frac{\M[\W_s]-f_s}{\tau_s},\qquad
            \partial_t\vb E=\partial_t\vb B=0,\\
 \Vop:&\quad \partial_tf_s+\kappa_s(\vb E+\vb u\times\vb B)\cdot\nabla_{\vb u}f_s=0,
            \qquad\text{coupled to \eqref{eq:maxwell}}.
 \end{aligned}
\end{equation*}
With $\delta=\Delta t/2$, one time step from $t^n$ to $t^{n+1}=t^n+\Delta t$ consists of four substeps
(Fig.~\ref{fig:step}): a transport step $\Aop_\delta$ of length $\delta$ (Section~\ref{sec:transport}), the field step
$\Vop_{\Delta t}$ of length $\Delta t$ (Section~\ref{sec:field}), a second transport step $\Aop_\delta$, and the Gauss
correction $\Gop$ (Section~\ref{sec:gauss}). The stages after the first three substeps are labeled $a$, $b$ and $c$.
\begin{figure}[!b]
\centering
\begin{tikzpicture}[font=\small,
  st/.style={draw,circle,minimum size=9.5mm,inner sep=0pt},
  op/.style={-{Stealth[length=2.2mm]},thick},
  wv/.style={font=\footnotesize,anchor=south},
  dsc/.style={font=\footnotesize,anchor=north}]
 \node[st] (s1) at (0,0) {$n$};
 \node[st] (s2) at (3.5,0) {$a$};
 \node[st] (s3) at (7.0,0) {$b$};
 \node[st] (s4) at (10.5,0) {$c$};
 \node[st] (s5) at (14.0,0) {$n\!+\!1$};
 \draw[op] (s1) -- node[above]{$\Aop_\delta$} (s2);
 \draw[op] (s2) -- node[above]{$\Vop_{\Delta t}$} (s3);
 \draw[op] (s3) -- node[above]{$\Aop_\delta$} (s4);
 \draw[op] (s4) -- node[above]{$\Gop$} (s5);
 \node[dsc] at (1.75,-0.55) {transport};
 \node[dsc] at (5.25,-0.55) {field};
 \node[dsc] at (8.75,-0.55) {transport};
 \node[dsc] at (12.25,-0.55) {Gauss correction};
 \node[wv] at (0,0.55) {$h^n_s$};
 \node[wv] at (3.5,0.55) {$\M[\W^{h,a}_s]$};
 \node[wv] at (7.0,0.55) {$h^b_s$};
 \node[wv] at (10.5,0.55) {$\M[\W^{h,c}_s]$};
 \node[wv] at (14.0,0.55) {$(1+\zeta_s)h^c_s$};
\end{tikzpicture}
\caption{Substeps of one time step of UGKWP-EC, with the wave of a charged species at each stage. The field step maps the
reconstructed wave $h^a_s=\M[\W^{h,a}_s]$ to the field-pushed wave $h^b_s$, from which the second transport step starts.
The Gauss correction scales the reconstructed wave $h^c_s=\M[\W^{h,c}_s]$ to the corrected wave $(1+\zeta_s)h^c_s$ of
\eqref{eq:gauss-update}, which is the wave $h^n_s$ of the next step.}
\label{fig:step}
\end{figure}

At every stage, the discrete state of species $s$ consists of its particles $P_k(m_k,\vb x_k,\vb u_k)$, its cell
averages $\W_{s,i}$ and its hydrodynamic wave $h_s(\vb x,\vb u)$ with the cell-averaged moments $\W^h_{s,i}$. The
electric and magnetic fields are represented by the discrete fields $\vb E_h$ and $\vb B_h$ (Section~\ref{sec:fe}). The particles of species $s$ in $\Omega_i$ form the set $\mathcal P_{s,i}$, and those in all cells form $\mathcal P_s$.
Their cell-averaged moments are $\W^p_{s,i}=|\Omega_i|^{-1}\sum_{k\in\mathcal P_{s,i}}m_k\vb\psi(\vb u_k)$.

In the UGKWP method of Section~\ref{sec:ugkwp}, the hydrodynamic wave $h_s(\vb x,\cdot)=\M[\W^h_s(\vb x)]$ is the
Maxwellian of the second-order reconstruction $\W^h_s(\vb x)$ from the cell averages $\W^h_{s,i}$. In UGKWP-EC, the field step
transforms the reconstructed wave $h^a_s$ of a charged species into the field-pushed wave $h^b_s$
\eqref{eq:pushed-wave} with the velocity map that also pushes the particles, and the Gauss correction scales the
reconstructed wave $h^c_s$ to $h^{n+1}_s$ \eqref{eq:gauss-update} with the factor that also multiplies the particle masses. These waves are sampled and transported by the
subsequent transport steps without being rebuilt from the averages $\W^h_{s,i}$.

\subsection{Transport step}\label{sec:transport}

The transport step $\Aop_\delta$ advances every species over the time $\delta$ by the UGKWP step of
Section~\ref{sec:ugkwp} for the BGK equation \eqref{eq:bgk}, with the relaxation time $\tau_s$ of the species. In this
subsection, the superscripts $n$ and $n+1$ refer to the beginning and the end of a transport step, and $\tau_{s,i}$ is
evaluated at $\W^n_{s,i}$.

The fluxes of species $s$ through an interface $S_{ij}$ take the coefficients $q_1$--$q_5$ of the upwind cell
\cite{Cao2026cf}, with the relaxation time $\tau_{s,i}$ of $\Omega_i$ for $u_n=\vb u\cdot\vb n_{ij}>0$ and $\tau_{s,j}$ of
$\Omega_j$ for $u_n<0$. The equilibrium flux is
\begin{equation*}
 \begin{aligned}
 \Feq_{s,ij}={}&\int_{u_n>0}u_n\vb\psi\,\bigl[q_1(\tau_{s,i})\,g_{s,0}+q_2(\tau_{s,i})\,\vb u\cdot\nabla g_{s,0}
 +q_3(\tau_{s,i})\,g_{s,t}\bigr]\dd\vb u\\
 &+\int_{u_n<0}u_n\vb\psi\,\bigl[q_1(\tau_{s,j})\,g_{s,0}+q_2(\tau_{s,j})\,\vb u\cdot\nabla g_{s,0}
 +q_3(\tau_{s,j})\,g_{s,t}\bigr]\dd\vb u,
 \end{aligned}
\end{equation*}
where the interface equilibrium $g_{s,0}$ and its derivatives $\nabla g_{s,0}$ and $g_{s,t}$ are formed from the
reconstruction of $\W_s$ as in Section~\ref{sec:ugkwp-fv}. As in \eqref{eq:ugkwp-frwave}, the free-transport flux of the
unsampled part of the wave is
\begin{equation*}
 \begin{aligned}
 \Ffw_{s,ij}={}&\int_{u_n>0}u_n\vb\psi\,\Bigl[\bigl(q_4(\tau_{s,i})-\delta p_{s,i}\bigr)h_{s,L}
 +\Bigl(q_5(\tau_{s,i})+\frac{\delta^2}{2}p_{s,i}\Bigr)\vb u\cdot\nabla h_{s,L}\Bigr]\dd\vb u\\
 &+\int_{u_n<0}u_n\vb\psi\,\Bigl[\bigl(q_4(\tau_{s,j})-\delta p_{s,j}\bigr)h_{s,R}
 +\Bigl(q_5(\tau_{s,j})+\frac{\delta^2}{2}p_{s,j}\Bigr)\vb u\cdot\nabla h_{s,R}\Bigr]\dd\vb u,
 \end{aligned}
\end{equation*}
where $h_{s,L}$, $\nabla h_{s,L}$ and $h_{s,R}$, $\nabla h_{s,R}$ are the wave $h^n_s$ and its gradient at $\vb x_{ij}$ in
$\Omega_i$ and $\Omega_j$, and $p_{s,i}=e^{-\delta/\tau_{s,i}}$ is the fraction of the wave sampled into particles in $\Omega_i$,
as in \eqref{eq:ugkwp-whp}.

\label{sec:update}As in \eqref{eq:ugkwp-update}, the averages are updated in conservation form,
\begin{equation}
 \W^{n+1}_{s,i}=\W^n_{s,i}-\frac1{|\Omega_i|}\sum_{j\in N(i)}
 \bigl(\Feq_{s,ij}+\Ffw_{s,ij}+\Ffp_{s,ij}\bigr)|S_{ij}|,
 \label{eq:fv-update}
\end{equation}
where the particles of species $s$ carry the net moments $\Ffp_{s,ij}|S_{ij}|$ across $S_{ij}$ from $\Omega_i$ into
$\Omega_j$, so that $-\sum_{j\in N(i)}\Ffp_{s,ij}|S_{ij}|$ equals the net particle moments \eqref{eq:ugkwp-pflux}.

At the end of the step, the mass of the wave is $\rho^{h,n+1}_{s,i}=\rho^{n+1}_{s,i}-\rho^{p,n+1}_{s,i}$. The residual $\W^{n+1}_{s,i}-\W^{p,n+1}_{s,i}$ contains the sampling noise of the particle
moments, and the relative noise of its thermal energy is large where the particles carry most of the mass of the cell. The
temperature of the wave is therefore set to that of a reference state, the expected residual of a spatially uniform state. In such a
state with $\W^{h,n}_{s,i}=\W^n_{s,i}-\W^{p,n}_{s,i}$ at the beginning of the step, the averages do not change, and the particles carry the expected moments $e^{-\delta/\tau_{s,i}}\W^n_{s,i}$ at its end. The
residual $\W^{n+1}_{s,i}-e^{-\delta/\tau_{s,i}}\W^n_{s,i}=(1-e^{-\delta/\tau_{s,i}})\W^{n+1}_{s,i}$ then has the velocity and temperature of
$\W^{n+1}_{s,i}$. The reference state
$\widehat\W^{n+1}_{s,i}=(\rho^{h,n+1}_{s,i}/\rho^{n+1}_{s,i})\W^{n+1}_{s,i}$ is this residual scaled to the mass of the wave. Where the particle fraction varies in space, the momentum $\vb P^{n+1}_{s,i}-\vb P^{p,n+1}_{s,i}$ of the residual
differs from $\widehat{\vb P}^{n+1}_{s,i}$, because the analytic fluxes bring momentum into $\W_{s,i}$ but not into the particles. The momentum of the wave is therefore taken from the residual and weighted toward
$\widehat{\vb P}^{n+1}_{s,i}$ where the wave is small, and its energy corresponds to the temperature of
$\widehat\W^{n+1}_{s,i}$,
\begin{equation}
 \begin{aligned}
 \vb P^{h,n+1}_{s,i}&=(1-\varpi_{s,i})\bigl(\vb P^{n+1}_{s,i}-\vb P^{p,n+1}_{s,i}\bigr)+\varpi_{s,i}\widehat{\vb P}^{n+1}_{s,i},\\
 \Ecal^{h,n+1}_{s,i}&=\widehat\Ecal^{n+1}_{s,i}+\frac{1}{2\rho^{h,n+1}_{s,i}}\bigl(|\vb P^{h,n+1}_{s,i}|^2
 -|\widehat{\vb P}^{n+1}_{s,i}|^2\bigr),
 \end{aligned}
 \label{eq:closure}
\end{equation}
with $\varpi_{s,i}=N_{\rm c}/(N_{\rm c}+N^h_{s,i})$, where $N^h_{s,i}=|\Omega_i|\rho^{h,n+1}_{s,i}/m_{{\rm ref},s}$ is the mass
of the wave in units of a prescribed reference particle mass $m_{{\rm ref},s}$ and $N_{\rm c}>0$ is a constant. Where the mass of the wave amounts to only a few reference particles, $N^h_{s,i}\ll N_{\rm c}$, the residual momentum is
noisy, and $\varpi_{s,i}$ approaches one.

\subsection{Field step}\label{sec:field}

The field step $\Vop_{\Delta t}$ advances the electric and magnetic fields over the time $\Delta t$, together with the
particle velocities and the waves, and keeps the particle positions $\vb x^a_k$ and masses. The preceding transport step does
not change the fields, so the field step starts from those of $t^n$. At the beginning of the field step, the wave
$h^a_s(\vb x,\cdot)=\M[\W^{h,a}_s(\vb x)]$ is reconstructed from the cell averages $\W^{h,a}_{s,i}$, and its moments
$\W^{h,a}_s=(\rho^{h,a}_s,\vb P^{h,a}_s,\Ecal^{h,a}_s)^\tr$ vary linearly within each cell. Cell integrals over
the wave are evaluated by the two-point tensor-product Gauss rule $\Iq_i$ on $\Omega_i$.

\label{sec:fe}The electric field is continuous and piecewise multilinear, determined by its values at the nodes of the
mesh, and the magnetic field is piecewise constant, determined by its values in the cells. With the vectors $\ms e$ and
$\ms b$ of these values, $\vb E_h(\vb x)=\ms N_E(\vb x)\ms e$ and $\vb B_h(\vb x)=\ms N_B(\vb x)\ms b$. The columns of $\ms N_E$
are $\lambda_\ell\hat{\vb x}$, $\lambda_\ell\hat{\vb y}$ and $\lambda_\ell\hat{\vb z}$, where $\{\lambda_\ell\}$ is the continuous
multilinear nodal basis of the Cartesian mesh, with $\lambda_\ell\ge0$ and $\sum_\ell\lambda_\ell=1$, and $\ms\Lambda(\vb x)$ is the
row vector of the $\lambda_\ell(\vb x)$. The columns of $\ms N_B$ are $\hat{\vb x}$, $\hat{\vb y}$ and $\hat{\vb z}$ times the
indicator functions of the cells.
With the mass and curl matrices
\begin{equation*}
 \ms M_E=\int_\Omega\ms N_E^\tr\ms N_E\dd\vb x,\qquad
 \ms M_B=\int_\Omega\ms N_B^\tr\ms N_B\dd\vb x,\qquad
 \ms C=\int_\Omega\ms N_B^\tr\,(\nabla\times\ms N_E)\dd\vb x,
\end{equation*}
the semi-discrete field equations with periodic or perfectly conducting boundaries are
\begin{equation*}
 \epsilon_0\ms M_E\frac{\dd\ms e}{\dd t}=\mu_0^{-1}\ms C^\tr\ms b-\ms j,\qquad
 \ms M_B\frac{\dd\ms b}{\dd t}=-\ms C\ms e,\qquad
 \ms j=\int_\Omega\ms N_E^\tr\vb J\dd\vb x ,
\end{equation*}
where $\ms j$ is the discrete current.

\label{sec:response}The electric and magnetic fields and the particle velocities are advanced by the implicit midpoint rule,
with the rotation field
$\vb B^{\rm rot}_h(\vb x)=\ms N_B(\vb x)\ms b^{\rm rot}$ in place of $\vb B$ in the Lorentz force. The wave is advanced by the same
velocity map. The rotation field, defined in \eqref{eq:frozen-b} below, is a magnetic field computed at the beginning of the field step.
With $\vb B^{\rm rot}_h$ known, the midpoint velocities vary linearly with the electric field, so that the field step
reduces to a linear system, as in ECsim \cite{Lapenta2017}. For species $s$, let
\begin{equation*}
 \begin{gathered}
 L_s(\vb x)\vb z=\kappa_s\,\vb z\times\vb B^{\rm rot}_h(\vb x),\qquad
 \alpha_s(\vb x)=\bigl(\I-\delta L_s(\vb x)\bigr)^{-1},\\
 \alpha_s\vb z=\frac{\vb z+\vb z\times\vb t_s+\vb t_s(\vb t_s\cdot\vb z)}{1+|\vb t_s|^2},
 \qquad \vb t_s=\delta\kappa_s\vb B^{\rm rot}_h .
 \end{gathered}
\end{equation*}

With $\bar{\ms e}=\tfrac12(\ms e^n+\ms e^b)$, the velocity of a particle $k$ of species $s$ is advanced by the implicit midpoint
rule
\begin{equation}
 \vb u^b_k-\vb u^a_k=\Delta t\,\kappa_s\bigl(\ms N_k\bar{\ms e}+\bar{\vb u}_k\times\vb B^{\rm rot}_k\bigr),
 \qquad \bar{\vb u}_k=\tfrac12(\vb u^a_k+\vb u^b_k),
 \label{eq:mover}
\end{equation}
where $\ms N_k=\ms N_E(\vb x^a_k)$ and $\vb B^{\rm rot}_k=\vb B^{\rm rot}_h(\vb x^a_k)$. Solving it for the midpoint velocity gives
\begin{equation}
 \bar{\vb u}_k=\alpha_k\bigl(\vb u^a_k+\delta\kappa_s\ms N_k\bar{\ms e}\bigr),\qquad
 \vb u^b_k=2\bar{\vb u}_k-\vb u^a_k,
 \label{eq:mover-explicit}
\end{equation}
with $\alpha_k=\alpha_s(\vb x^a_k)$.
At fixed $\vb x$, the midpoint velocity that \eqref{eq:mover-explicit} assigns to a velocity $\vb u$ of $h^a_s(\vb x,\cdot)$ is
$\alpha_s(\vb x)\bigl(\vb u+\delta\kappa_s\ms N_E(\vb x)\bar{\ms e}\bigr)$. Its moment over
$h^a_s(\vb x,\cdot)$ is the midpoint momentum density of the wave,
\begin{equation*}
 \bar{\vb P}^h_s=\alpha_s\bigl(\vb P^{h,a}_s+\delta\kappa_s\rho^{h,a}_s\ms N_E\bar{\ms e}\bigr).
\end{equation*}
With the charge $m_k\kappa_s$ of particle $k$ and the cell quadrature $\Iq_i$ for the wave, the discrete current at
the midpoint is
\begin{equation*}
 \bar{\ms j}=\sum_s\kappa_s\Bigl(\sum_{k\in\mathcal P_s}m_k\ms N_k^\tr\bar{\vb u}_k
 +\sum_i\Iq_i\bigl[\ms N_E^\tr\bar{\vb P}^h_s\bigr]\Bigr).
\end{equation*}
By \eqref{eq:mover-explicit} and the expression for $\bar{\vb P}^h_s$, $\bar{\ms j}$ varies linearly
with $\bar{\ms e}$ through the particle and wave responses
\begin{equation}
 \hat{\ms j}_p=\sum_s\sum_{k\in\mathcal P_s}m_k\kappa_s\ms N_k^\tr\alpha_k\vb u^a_k,\qquad
 \ms X_p=\sum_s\sum_{k\in\mathcal P_s}m_k\kappa_s^2\ms N_k^\tr\alpha_k\ms N_k,
 \label{eq:resp-particle}
\end{equation}
\begin{equation}
 \begin{gathered}
 \hat{\ms j}_h=\sum_s\sum_i\Iq_i\bigl[\kappa_s\ms N_E^\tr\alpha_s\vb P^{h,a}_s\bigr],\qquad
 \ms X_h=\sum_s\sum_i\Iq_i\bigl[\kappa_s^2\rho^{h,a}_s\ms N_E^\tr\alpha_s\ms N_E\bigr],\\
 \hat{\ms j}=\hat{\ms j}_p+\hat{\ms j}_h,\qquad \ms X=\ms X_p+\ms X_h,\qquad
 \bar{\ms j}=\hat{\ms j}+\delta\ms X\bar{\ms e}.
 \end{gathered}
 \label{eq:resp-wave}
\end{equation}

\label{sec:mixed}The field update is the implicit midpoint rule for the semi-discrete field equations, with the
discrete current $\bar{\ms j}$,
\begin{equation}
 \begin{aligned}
 \epsilon_0\ms M_E\frac{\ms e^b-\ms e^n}{\Delta t}&=\mu_0^{-1}\ms C^\tr\bar{\ms b}-\bar{\ms j},\\
 \ms M_B\frac{\ms b^b-\ms b^n}{\Delta t}&=-\ms C\bar{\ms e},\qquad
 \bar{\ms b}=\tfrac12(\ms b^b+\ms b^n).
 \end{aligned}
 \label{eq:discrete-maxwell}
\end{equation}
The discrete Faraday law in \eqref{eq:discrete-maxwell} gives $\bar{\ms b}=\ms b^n-\delta\ms M_B^{-1}\ms C\bar{\ms e}$, and the
discrete Amp\`ere law, multiplied by $\delta$ and written in $\bar{\ms e}$ with \eqref{eq:resp-wave}, becomes a linear system for the midpoint electric field,
\begin{equation}
 \Bigl(\epsilon_0\ms M_E+\delta^2\ms X+\frac{\delta^2}{\mu_0}\ms C^\tr\ms M_B^{-1}\ms C\Bigr)\bar{\ms e}
 =\epsilon_0\ms M_E\ms e^n-\delta\hat{\ms j}+\frac{\delta}{\mu_0}\ms C^\tr\ms b^n .
 \label{eq:schur}
\end{equation}
Equation~\eqref{eq:schur} is an RME \cite{Degond2017a,Liu2025}, in which the current
response \eqref{eq:resp-wave} is the discrete form of the current equation
$\partial_t\vb J=\sum_s\kappa_s^2(\rho_s\vb E+\vb P_s\times\vb B)$ of the operator $\Vop$.

The RME has exactly one solution for every $\Delta t>0$ (Proposition~\ref{prop:energy}), and the
endpoint fields are
\begin{equation}
 \ms e^b=2\bar{\ms e}-\ms e^n,\qquad
 \ms b^b=2\bar{\ms b}-\ms b^n=\ms b^n-\Delta t\,\ms M_B^{-1}\ms C\bar{\ms e}.
 \label{eq:endpoint}
\end{equation}
Since the field step changes $\ms b$ by $-\Delta t\,\ms M_B^{-1}\ms C\bar{\ms e}$ in \eqref{eq:endpoint} and the other
substeps leave $\ms b$ unchanged, $\ms b^n-\ms b^0$ lies in the range of the discrete curl $\ms M_B^{-1}\ms C$. On the Cartesian meshes, the weak
divergence $\int_\Omega\nabla\lambda_\ell\cdot\vb B_h\dd\vb x$ vanishes on this range at every node $\ell$, as both operators are
composed of differences and averages along the coordinate directions, which commute. This weak divergence therefore keeps
its initial value, so that the method needs no divergence cleaning of the magnetic field.

\label{sec:push}With $\bar{\ms e}$ from \eqref{eq:schur}, every particle is pushed once by \eqref{eq:mover-explicit}. Written without
$\bar{\vb u}_k$, the push is the velocity map $\vb u^b_k=R_s(\vb x^a_k)\vb u^a_k+\vb d_s(\vb x^a_k)$ with
\begin{equation}
 R_s(\vb x)=2\alpha_s(\vb x)-\I,\qquad
 \vb d_s(\vb x)=\Delta t\,\kappa_s\,\alpha_s(\vb x)\ms N_E(\vb x)\bar{\ms e},
 \label{eq:map}
\end{equation}
where $R_s=(\I+\delta L_s)(\I-\delta L_s)^{-1}$ is orthogonal, as the Cayley transform of the skew-symmetric matrix
$\delta L_s$. The field-pushed wave is the image of $h^a_s(\vb x,\cdot)$ under the same map
$\vb u\mapsto R_s(\vb x)\vb u+\vb d_s(\vb x)$, which preserves volume in velocity space,
\begin{equation}
 h^b_s(\vb x,\vb u)=h^a_s\bigl(\vb x,\,R_s(\vb x)^\tr[\vb u-\vb d_s(\vb x)]\bigr).
 \label{eq:pushed-wave}
\end{equation}
The map moves the velocity $\vb U^{h,a}_s$ of the Maxwellian $h^a_s(\vb x,\cdot)$ to
$R_s\vb U^{h,a}_s+\vb d_s$ and keeps its temperature, as $R_s$ is orthogonal. Hence
$h^b_s(\vb x,\cdot)=\M[\W^{h,b}_s(\vb x)]$, with
\begin{equation}
 \begin{gathered}
 \rho^{h,b}_s=\rho^{h,a}_s,\qquad
 \vb U^{h,b}_s=R_s\vb U^{h,a}_s+\vb d_s,\qquad
 T^{h,b}_s=T^{h,a}_s,\\
 \vb P^{h,b}_s=R_s\vb P^{h,a}_s+\rho^{h,a}_s\vb d_s,\qquad
 \Ecal^{h,b}_s=\Ecal^{h,a}_s+\vb d_s\cdot R_s\vb P^{h,a}_s+\tfrac12\rho^{h,a}_s|\vb d_s|^2.
 \end{gathered}
 \label{eq:pushed-moments}
\end{equation}
The cell averages are updated by the increments of the particles and of the wave,
\begin{equation}
 \begin{gathered}
 \Delta\W^p_{s,i}=\frac1{|\Omega_i|}\sum_{k\in\mathcal P_{s,i}}m_k\bigl[\vb\psi(\vb u^b_k)-\vb\psi(\vb u^a_k)\bigr],\qquad
 \Delta\W^h_{s,i}=\frac1{|\Omega_i|}\Iq_i\bigl[\W^{h,b}_s-\W^{h,a}_s\bigr],\\
 \W^b_{s,i}=\W^a_{s,i}+\Delta\W^p_{s,i}+\Delta\W^h_{s,i},\qquad
 \W^{h,b}_{s,i}=\W^{h,a}_{s,i}+\Delta\W^h_{s,i}.
 \end{gathered}
 \label{eq:material-push}
\end{equation}
As the moments \eqref{eq:pushed-moments} are polynomials of degree at most three in each coordinate, $\Iq_i$
integrates them exactly, and $\W^{h,b}_{s,i}$ is the cell average of the moments of $h^b_s$.

\label{sec:properties}The total energy of a state with the cell averages $\W_{s,i}=(\rho_{s,i},\vb P_{s,i},\Ecal_{s,i})^\tr$ and
the field values $\ms e$ and $\ms b$ is
\[
 \Ecal_{\rm tot}=\sum_{s,i}|\Omega_i|\,\Ecal_{s,i}+\frac{\epsilon_0}{2}\,\ms e^\tr\ms M_E\ms e
 +\frac{1}{2\mu_0}\,\ms b^\tr\ms M_B\ms b .
\]

\begin{proposition}\label{prop:energy}
Let the rotation field $\ms b^{\rm rot}$ be arbitrary. Then:
\begin{enumerate}
\item[(i)] Equation \eqref{eq:schur} has exactly one solution $\bar{\ms e}$ for every $\Delta t>0$.
\item[(ii)] The particle and wave updates \eqref{eq:mover-explicit} and \eqref{eq:map}--\eqref{eq:material-push} change
the energy of the species by
$\sum_{s,i}|\Omega_i|\bigl(\Ecal^b_{s,i}-\Ecal^a_{s,i}\bigr)=\Delta t\,\bar{\ms e}^\tr\bar{\ms j}$, with $\bar{\ms j}$ of
\eqref{eq:resp-wave}.
\item[(iii)] With periodic or perfectly conducting field boundaries, $(\W^a,\ms e^n,\ms b^n)$ and
$(\W^b,\ms e^b,\ms b^b)$ have the same total energy.
\end{enumerate}
\end{proposition}

\begin{proof}
(i) As $L_s$ is skew-symmetric, $\alpha_s^{-1}+\alpha_s^{-\tr}=2\I$ and $\vb z^\tr\alpha_s\vb z=|\alpha_s\vb z|^2$. By
\eqref{eq:resp-particle}, \eqref{eq:resp-wave} and $\rho^{h,a}_s\ge0$, therefore,
\[
 \ms w^\tr\ms X\ms w=\sum_s\kappa_s^2\Bigl(\sum_{k\in\mathcal P_s}m_k\bigl|\alpha_k\ms N_k\ms w\bigr|^2
 +\sum_i\Iq_i\bigl[\rho^{h,a}_s\bigl|\alpha_s\ms N_E\ms w\bigr|^2\bigr]\Bigr)\ge0
\]
for every $\ms w$. As $\ms M_E$ is positive definite and $\ms C^\tr\ms M_B^{-1}\ms C$ is positive semidefinite, the
quadratic form of the matrix of \eqref{eq:schur} is at least $\epsilon_0\ms w^\tr\ms M_E\ms w>0$ for $\ms w\ne0$, and the
matrix is nonsingular.

(ii) For a particle, \eqref{eq:mover}, \eqref{eq:mover-explicit} and $\bar{\vb u}_k\cdot(\bar{\vb u}_k\times\vb B^{\rm rot}_k)=0$ give
\[
 \tfrac12\bigl(|\vb u^b_k|^2-|\vb u^a_k|^2\bigr)=\bar{\vb u}_k\cdot(\vb u^b_k-\vb u^a_k)
 =\Delta t\,\kappa_s\,\bar{\ms e}^\tr\ms N_k^\tr\alpha_k\bigl(\vb u^a_k+\delta\kappa_s\ms N_k\bar{\ms e}\bigr),
\]
whose sum with the weights $m_k$ is $\Delta t\,\bar{\ms e}^\tr(\hat{\ms j}_p+\delta\ms X_p\bar{\ms e})$ by
\eqref{eq:resp-particle}. The wave is pushed by the same velocity map \eqref{eq:map} as the particles, so the identity holds at
every point $\vb x$ for every velocity $\vb u$. Its integral against $h^a_s(\vb x,\vb u)\dd\vb u$ is, by \eqref{eq:pushed-wave},
$\Ecal^{h,b}_s-\Ecal^{h,a}_s=\Delta t\,\kappa_s\bar{\ms e}^\tr\ms N_E^\tr\bar{\vb P}^h_s$, whose quadratures in
\eqref{eq:material-push} sum by \eqref{eq:resp-wave} to $\Delta t\,\bar{\ms e}^\tr(\hat{\ms j}_h+\delta\ms X_h\bar{\ms e})$.

(iii) The fields \eqref{eq:endpoint} satisfy \eqref{eq:discrete-maxwell} with
$\bar{\ms b}=\frac12(\ms b^b+\ms b^n)=\ms b^n-\delta\ms M_B^{-1}\ms C\bar{\ms e}$.
As $\ms M_E$ and $\ms M_B$ are symmetric and $\ms M_B(\ms b^b-\ms b^n)=-\Delta t\,\ms C\bar{\ms e}$, the field
energy changes by $\epsilon_0\bar{\ms e}^\tr\ms M_E(\ms e^b-\ms e^n)+\mu_0^{-1}\bar{\ms b}^\tr\ms M_B
(\ms b^b-\ms b^n)=\Delta t\,\mu_0^{-1}(\bar{\ms e}^\tr\ms C^\tr\bar{\ms b}-\bar{\ms b}^\tr\ms C\bar{\ms e})
-\Delta t\,\bar{\ms e}^\tr\bar{\ms j}=-\Delta t\,\bar{\ms e}^\tr\bar{\ms j}$, and (ii) gives (iii).
\end{proof}

\label{sec:rotation}Proposition~\ref{prop:energy} holds for any rotation field, so that the energy balance does not
constrain its choice. The implicit midpoint rule evaluates the Lorentz force at the midpoint $t^n+\delta$ of the field step. The method predicts the
magnetic field at $t^n+\delta$ by Faraday's law from the filtered electric field $\ms S^2\ms e^n$, and reduces the
predicted change where $\omega_{p,i}\Delta t$ or the drift parameter $\beta_i$ defined below is large,
\begin{equation}
 \ms b^{\rm rot}=\ms b^n-\delta\,\ms\Phi\,\ms M_B^{-1}\ms C\,\ms S^2\ms e^n,
 \label{eq:frozen-b}
\end{equation}
where $\ms S$ averages the nodal values of $\ms e$ with the weights $(\frac14,\frac12,\frac14)$ in each
coordinate direction and thus removes the grid-scale oscillation of $\ms e^n$ from $\ms b^{\rm rot}$. The diagonal matrix
$\ms\Phi$ multiplies the values in each cell $\Omega_i$ by the factor
\[
 \begin{gathered}
 \Phi_i=\Bigl(1+\frac{(\omega_{p,i}\Delta t)^2}{48}+\frac{\beta_i^2}{16}\Bigr)^{-1},\qquad
 \beta_i=k_*V_i\,\omega_{p,i}\Delta t^2,\qquad k_*=\frac2{h_{\min}},\\
 \omega_{p,i}^2=\frac1{\epsilon_0}\sum_s\kappa_s^2\rho^a_{s,i},\qquad
 \omega_{p,i}^2V_i^2=\frac1{\epsilon_0}\sum_s\kappa_s^2\frac{|\vb P^a_{s,i}|^2}{\rho^a_{s,i}}.
 \end{gathered}
\]
Here $\omega_{p,i}$ is the plasma frequency, $V_i$ the root-mean-square drift speed of the charged species in $\Omega_i$,
$h_{\min}$ the smallest cell width and $k_*$ the largest modified wavenumber of the discrete curl. Without the drift term, $\Phi_i$ damps the predicted change as a time filter of width $\Delta t$ damps an
oscillation at the plasma frequency. It agrees to second order in $\omega_{p,i}\Delta t$ with
$\operatorname{sinc}^2(\omega_{p,i}\Delta t/4)$, the transfer function of a triangular filter. It decays as $(\omega_{p,i}\Delta t)^{-2}$, so that
$\ms b^{\rm rot}$ tends to $\ms b^n$ where the time step does not resolve the plasma period.
The drift term reaches one at $\beta_i=4$. Below this bound, the time step with $\ms\Phi$ and $\ms S$ omitted in \eqref{eq:frozen-b}
keeps a positive definite quadratic invariant that bounds the perturbations of a drifting cold plasma (\ref{app:drift}).

\subsection{Gauss correction}\label{sec:gauss}

The Gauss correction $\Gop$ removes the
part of the nodal Gauss defect that lies in the range of the weak divergence, with the least change of the electric
field and of the masses of the particles and the wave, as measured by \eqref{eq:gauss-cost}. It leaves the positions and velocities
of the particles and the magnetic field unchanged.

With the moments $\W^{h,c}_s(\vb x)=(\rho^{h,c}_s,\vb P^{h,c}_s,\Ecal^{h,c}_s)^\tr$ of the reconstructed wave
$h^c_s$, the nodal charges before the correction and the weak divergence are
\begin{equation}
 \ms Q=\sum_s\kappa_s\Bigl(\sum_{k\in\mathcal P_s}m_k\ms\Lambda(\vb x_k)^\tr+\int_\Omega\ms\Lambda^\tr\rho^{h,c}_s\dd\vb x\Bigr),
 \qquad
 \ms D=\int_\Omega\ms\Lambda^\tr\,(\nabla\cdot\ms N_E)\dd\vb x,
 \label{eq:gauss-defect}
\end{equation}
and the Gauss defect is $\ms r=\epsilon_0\ms D\ms e^c-\ms Q$. Its component $\ms r_\ell$ is the residual of Gauss's law
$\epsilon_0\nabla\cdot\vb E_h=\rho_c$ in weak form with the test function $\lambda_\ell$, where the particles contribute to $\rho_c$ as point
charges. A change $\Delta\ms e$ of the electric field changes $\ms r$ by $\epsilon_0\ms D\Delta\ms e$, that is, only within the
range of $\ms D$. The correction $\Gop$ therefore removes the part $\ms r_\parallel$ of the defect that lies in the range of
$\ms D$. The remaining part $\ms r_\perp=\ms r-\ms r_\parallel$ satisfies $\ms D^\tr\ms r_\perp=0$.

For a charged species $s$ and a cell $\Omega_i$ that it occupies ($\W^c_{s,i}\ne0$), the correction multiplies the
masses of the particles of $\mathcal P_{s,i}$ and the wave of $s$ on $\Omega_i$ by the same function $1+\zeta_s$,
\begin{equation}
 \zeta_s(\vb x)=\sum_\ell\zeta_{s,i,\ell}\lambda_\ell(\vb x)\quad(\vb x\in\Omega_i),\qquad
 \zeta_{s,i,\ell}\ge-1,
 \label{eq:gauss-multiplier}
\end{equation}
where the sum runs over the nodes $\ell$ of $\Omega_i$. As $\lambda_\ell\ge0$ and $\sum_\ell\lambda_\ell=1$, the bounds give
$1+\zeta_s\ge0$, so that the masses of the particles and the wave stay nonnegative. The corrected state is
\begin{equation}
 \begin{gathered}
 m^{n+1}_k=\bigl(1+\zeta_s(\vb x_k)\bigr)m_k,\qquad
 h^{n+1}_s(\vb x,\vb u)=\bigl(1+\zeta_s(\vb x)\bigr)h^c_s(\vb x,\vb u),\\[2pt]
 \W^{n+1}_{s,i}=\W^c_{s,i}+\frac1{|\Omega_i|}\Bigl(\sum_{k\in\mathcal P_{s,i}}m_k\zeta_s(\vb x_k)\vb\psi(\vb u_k)
 +\int_{\Omega_i}\zeta_s\W^{h,c}_s\dd\vb x\Bigr),\\[2pt]
 \W^{h,n+1}_{s,i}=\W^{h,c}_{s,i}+\frac1{|\Omega_i|}\int_{\Omega_i}\zeta_s\W^{h,c}_s\dd\vb x,
 \end{gathered}
 \label{eq:gauss-update}
\end{equation}
with the electric field $\ms e^{n+1}$ determined below. The corrected charges
$\ms Q^{n+1}$ follow from \eqref{eq:gauss-defect} with $m^{n+1}_k$ and $(1+\zeta_s)\rho^{h,c}_s$ in place of $m_k$ and
$\rho^{h,c}_s$, and
$\ms Q^{n+1}$, $\W^{n+1}_{s,i}$ and $\W^{h,n+1}_{s,i}$ vary linearly with the coefficients.

The field $\ms e^{n+1}$ and the coefficients $\zeta_{s,i,\ell}$ minimize
\begin{equation}
 \mathcal J=\frac{\epsilon_0}{2}(\ms e^{n+1}-\ms e^c)^\tr\ms M_E(\ms e^{n+1}-\ms e^c)
 +\frac12\sum_{s,i}\frac{k_BT_{s,i}}{m_s}\Bigl(\sum_{k\in\mathcal P_{s,i}}m_k\zeta_s(\vb x_k)^2
 +\int_{\Omega_i}\rho^{h,c}_s\zeta_s^2\dd\vb x\Bigr)
 \label{eq:gauss-cost}
\end{equation}
subject to the bounds in \eqref{eq:gauss-multiplier} and to
\begin{equation}
 \epsilon_0\ms D\ms e^{n+1}-\ms Q^{n+1}=\ms r_\perp,\qquad
 \sum_i|\Omega_i|\bigl(\W^{n+1}_{s,i}-\W^c_{s,i}\bigr)=0\quad(\kappa_s\ne0),
 \label{eq:gauss-constraints}
\end{equation}
so that the Gauss defect after the correction is $\ms r_\perp$ and the total mass, momentum and energy of every charged
species are conserved. The first term of $\mathcal J$ is the electric energy of the field increment. The second term is the quadratic part of the change of the isothermal free energy
$(k_BT_{s,i}/m_s)\int\rho\ln\rho\dd\vb x$ when the mass density $\rho$ of the particles and the wave is multiplied by
$1+\zeta_s$.

For a uniform plasma on a periodic domain, let $r=\epsilon_0\nabla\cdot\vb E-\rho_c$ be a small, smooth defect with zero
mean. The continuous form of the minimization is then solved by $\Delta\vb E=-\nabla\phi$ and
$\zeta_s=-\kappa_sm_s\phi/(k_BT_s)$, where $-\nabla^2\phi+\lambda_D^{-2}\phi=-r/\epsilon_0$ and $\lambda_D$ is the Debye length.
Since $\phi$ has zero mean, this solution also satisfies the
conservation constraints. It is linear Debye screening of the charge $-r$, which removes the defect mainly through the
charge on scales larger than the Debye length. The smallest correction of the field alone solves the same equation without
the term $\lambda_D^{-2}\phi$. As $\epsilon_0\to0$ for a given plasma and defect, its field change grows as $1/\epsilon_0$,
whereas that of this solution stays bounded. This is why the correction also adjusts the masses: in quasi-neutral regions, a
correction of the field alone turns small charge errors into large electric fields. Beyond this model, the electric energy of the field increment of
$\Gop$ never exceeds that of a correction of the field alone, since $\zeta=0$ with such a field
satisfies all constraints and the second term of $\mathcal J$ is nonnegative.

The correction has the following properties, proved in \ref{app:proofs}.
\begin{proposition}\label{prop:gauss}\leavevmode
\begin{enumerate}
\item[(i)] The minimization of \eqref{eq:gauss-cost} subject to \eqref{eq:gauss-multiplier} and
\eqref{eq:gauss-constraints} has a solution, and all solutions give the same corrected state \eqref{eq:gauss-update}.
\item[(ii)] In this state, the Gauss defect $\epsilon_0\ms D\ms e^{n+1}-\ms Q^{n+1}$ is $\ms r_\perp$, and the total
mass, momentum and energy of every charged species are those before the correction.
\item[(iii)] The total energy and the electric energy $\frac{\epsilon_0}{2}\ms e^\tr\ms M_E\ms e$ change by the same amount.
\end{enumerate}
\end{proposition}

The minimization is a convex quadratic problem whose only inequality constraints are the bounds in
\eqref{eq:gauss-multiplier}. It is solved by the primal--dual active-set method \cite{Hintermueller2002}. Each
iteration eliminates the coefficients cell by cell and solves one linear system for $\ms e^{n+1}$ and the Lagrange
multipliers of \eqref{eq:gauss-constraints}.

\subsection{Algorithm}\label{sec:algorithm}

One time step of the method from $t^n$ to $t^{n+1}$ consists of the following steps (Fig.~\ref{fig:step}). Steps~1
and 3 are the UGKWP step of Section~\ref{sec:ugkwp-algorithm} as modified in Section~\ref{sec:transport}.
\begin{description}
\item[Step 1.] Perform the first transport step $\Aop_\delta$, starting from the wave $h^n_s$, with $\tau_{s,i}$
evaluated at $\W^n_{s,i}$. It gives the averages $\W^a_{s,i}$ and $\W^{h,a}_{s,i}$.
\item[Step 2.] Perform the field step $\Vop_{\Delta t}$. Reconstruct the wave $h^a_s$ from
the averages $\W^{h,a}_{s,i}$, form the rotation field \eqref{eq:frozen-b}, assemble the responses \eqref{eq:resp-particle}, \eqref{eq:resp-wave}, solve the RME \eqref{eq:schur}
for $\bar{\ms e}$, and set $\ms e^b$ and $\ms b^b$ \eqref{eq:endpoint}. Push every particle once \eqref{eq:mover-explicit},
form the field-pushed wave $h^b_s$ \eqref{eq:pushed-wave}, and update $\W^b_{s,i}$ and $\W^{h,b}_{s,i}$
\eqref{eq:material-push}.
\item[Step 3.] Perform the second transport step $\Aop_\delta$ as in Step~1, starting from the state after Step~2,
with $\tau_{s,i}$ evaluated at $\W^b_{s,i}$ and the field-pushed wave $h^b_s$
in place of $h^n_s$.
\item[Step 4.] Apply the Gauss correction $\Gop$. Reconstruct the wave $h^c_s$ from the averages $\W^{h,c}_{s,i}$,
form the defect \eqref{eq:gauss-defect}, solve \eqref{eq:gauss-cost}--\eqref{eq:gauss-constraints} by the active-set
iteration of Section~\ref{sec:gauss}, and set $\ms e^{n+1}$, the particle masses $m^{n+1}_k$, the wave
$h^{n+1}_s=(1+\zeta_s)h^c_s$, $\W^{n+1}_{s,i}$ and $\W^{h,n+1}_{s,i}$ \eqref{eq:gauss-update}, with $\ms b^{n+1}=\ms b^b$.
\end{description}

With periodic boundaries, the transport steps change neither $\ms e$ nor $\ms b$ and conserve the total energy, as they
update the averages in the conservation form \eqref{eq:fv-update}. By
Propositions~\ref{prop:energy}(iii) and \ref{prop:gauss}(iii), the total energy therefore changes over a complete time
step by
\[
 \Ecal^{n+1}_{\rm tot}-\Ecal^n_{\rm tot}=\frac{\epsilon_0}{2}\bigl[(\ms e^{n+1})^\tr\ms M_E\ms e^{n+1}
 -(\ms e^c)^\tr\ms M_E\ms e^c\bigr],
\]
which is the change of the electric energy in the Gauss correction.

\section{Numerical tests}\label{sec:tests}

All quantities are nondimensional in the sense of Section~\ref{sec:model-units}. The electron plasma and cyclotron
frequencies of the initial state are $\omega_{pe}=(\kappa_e^2\rho_e/\epsilon_0)^{1/2}$ and $\Omega_{ce}=|\kappa_e||\vb B|$.
In one-dimensional problems, all quantities depend only on the coordinate $y$, and the domain is divided into cells of
width $\Delta y$. The gradients of the reconstruction are limited by the van Leer limiter in each coordinate direction.

The time step is determined by the CFL condition on the flow and sound speeds of the species,
\begin{equation}
 \Delta t=\mathrm{CFL}\,\frac{h_{\min}}{\max_{s,i}\bigl(|\vb U_{s,i}|+c_{s,i}\bigr)},\qquad
 c_{s,i}=\sqrt{\frac{5k_BT_{s,i}}{3m_s}},
 \label{eq:tests-cfl}
\end{equation}
where $c_{s,i}$ is the sound speed of the monatomic species and $|\vb U_{s,i}|$ the flow speed. In all tests of this section,
UGKWP-EC uses $\mathrm{CFL}=0.8$.

In one dimension, the RME \eqref{eq:schur} and the linear systems of the Gauss correction are solved by LU factorization. In two
dimensions, the RME is solved by GMRES \cite{Saad1986} with a restricted additive Schwarz preconditioner \cite{Cai1999}, and the linear systems of the Gauss correction by
MINRES \cite{Paige1975} with a Chebyshev polynomial preconditioner. Both are solved to the relative residual $10^{-8}$
in Section~\ref{sec:tests-ot}, and in Section~\ref{sec:tests-gem} the RME to $10^{-10}$ and the Gauss correction to $10^{-6}$. Over a whole computation, the residual of the RME changes the total energy by less than
$6\times10^{-12}$ of its value.

The comparison method UGKWP-PHM is the UGKWP method of \cite{Pu2025a} with the same collision model. It advances the
perfectly hyperbolic Maxwell (PHM) equations \cite{Munz2000} explicitly by a second-order finite-volume scheme with the
monotonized central (MC) limiter. The damping rate of the cleaning fields is $\sigma_\phi=0$ in the periodic one-dimensional problems and $\sigma_\phi=3$ in the Brio--Wu problem and in two dimensions, where the computations
require the damping for stability. The time step of UGKWP-PHM is \eqref{eq:tests-cfl} with $\max(|\vb U_{s,i}|+c_{s,i},c)$ in place of
$|\vb U_{s,i}|+c_{s,i}$, at $\mathrm{CFL}=0.5$ in one dimension and $0.4$ in two dimensions. The two-dimensional problems are also
computed by UGKWP-PHM at $\mathrm{CFL}=0.5$, the stability limit of its Maxwell solver.

The CPU times reported in Sections~\ref{sec:tests-ot} and \ref{sec:tests-gem} are wall-clock times multiplied by the number of
cores. They are measured on 64 cores of a workstation with two 64-core AMD EPYC 7742 processors.

\subsection{Plasma waves}\label{sec:tests-waves}

Plasma waves test the longitudinal and the magnetized transverse response of the field step and its energy balance. An electron plasma wave covers the transition from collisionless Landau damping to collision-dominated oscillations, and a parallel whistler wave in a warm plasma tests the frequency shift from the cold-plasma value and the electron
cyclotron damping. With a collision
frequency that varies over four orders of magnitude in space, the electron plasma wave also tests the method at the transitions
between weakly collisional and collision-dominated regions.

The electron plasma wave is computed with $m_e=1$, $\kappa_e=-1$, $\epsilon_0=1$ and $c=20$. On the periodic domain
$[0,2\pi]$, the initial electron distribution function is the Maxwellian with zero mean velocity, the temperature
$T_e=0.25$ and the density
\[
 n_e(y,0)=1-ak\cos ky,\qquad k=1,\quad a=0.05,
\]
so that $\omega_{pe}=1$ and $k\lambda_{De}=0.5$, with the electron Debye length $\lambda_{De}=(T_e/m_e)^{1/2}/\omega_{pe}$. The ions are uniform and collisionless, with $m_i=1836$, $\kappa_i=1/1836$
and $T_i=0.01$, and the electron collision frequency is $\nu_e=\chi_{ee}n_e$ with $\chi_{ee}=0$, $1$ and $10$, that is,
$\nu_e=0$, $\omega_{pe}$ and $10\omega_{pe}$. For $\nu_e=0$, the electron particles are placed at $y_0+a\sin ky_0$ with uniformly spaced $y_0$. The initial electric field $E_y$ satisfies Gauss's law, and $\vb B=\vb 0$. The mesh has 64 cells, and the reference number of particles per cell is $4\times10^4$ for each species. The frequency $\omega$ and the damping rate $\gamma$ are obtained by fitting $A_\infty+A_1e^{\gamma t}\cos(\omega t+\varphi)$ to the coefficient $A(t)$ of $\sin ky$ in $E_y$
over $2\le t\le12$, where $A_\infty$ is the static field of the entropy mode. The reference, DVM-Poisson, solves the BGK--Vlasov--Poisson system with the same ions, $\nu_e$
and initial state by an independent discrete-velocity method (DVM) on 256 cells, with 800 discrete velocity points in $[-4.5,4.5]$ for the electrons and 128 for the ions.

\begin{figure}[t]
\centering
\includegraphics[width=\textwidth]{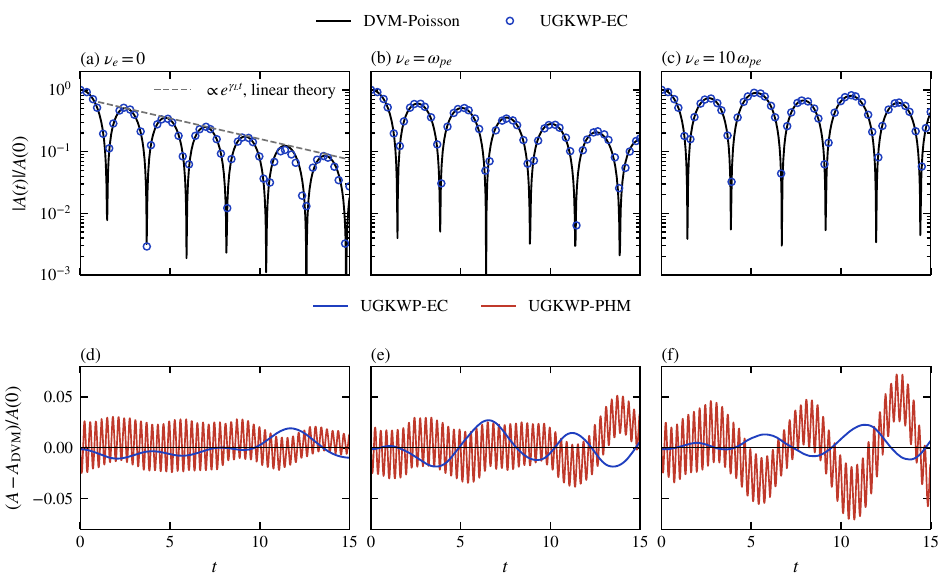}
\caption{Electron plasma wave with (a), (d) $\nu_e=0$, (b), (e) $\nu_e=\omega_{pe}$ and (c), (f) $\nu_e=10\omega_{pe}$. (a)--(c) $|A(t)|/A(0)$ of UGKWP-EC and of
the DVM-Poisson reference, with a line proportional to $e^{\gamma_Lt}$ in (a), where $A(t)$ is the coefficient of $\sin ky$ in $E_y$ and $\gamma_L=-0.1533$ is the linear Landau damping rate. (d)--(f) Difference of $A(t)$ of UGKWP-EC and
UGKWP-PHM from the DVM-Poisson reference, normalized by $A(0)$.}
\label{fig:tests-langmuir}
\end{figure}

\begin{table}[b]
\centering
\caption{Plasma waves: frequency $\omega$ and damping rate $\gamma$ of the reference and of UGKWP-EC. The reference is
DVM-Poisson for the electron plasma wave and the root of the warm kinetic dispersion relation for the whistler wave.}
\label{tab:tests-waves}
\footnotesize
\setlength{\tabcolsep}{4pt}
\begin{tabular*}{\textwidth}{@{\extracolsep{\fill}}lc d{1.5}d{1.5} d{2.5}d{2.5}@{}}
\toprule
 & & \multicolumn{2}{c}{$\omega$} & \multicolumn{2}{c}{$\gamma$}\\
\cmidrule(lr){3-4}\cmidrule(lr){5-6}
Wave & $\nu_e/\omega_{pe}$ & \colhead{Reference} & \colhead{UGKWP-EC} & \colhead{Reference} & \colhead{UGKWP-EC}\\
\midrule
Electron plasma & $0$ & 1.4137 & 1.4128 & -0.1584 & -0.1674\\
 & $1$ & 1.2591 & 1.2564 & -0.1065 & -0.1134\\
 & $10$ & 1.1918 & 1.1903 & -0.0186 & -0.0166\\
Whistler & $0$ & 0.12500 & 0.12568 & -0.02132 & -0.02166\\
\bottomrule
\end{tabular*}
\end{table}

As $\nu_e$ increases, the collisionless Landau damping gives way to a weakly damped oscillation whose frequency
approaches that of an adiabatic electron fluid, $(1+\frac53k^2\lambda_{De}^2)^{1/2}\omega_{pe}=1.190$
(Fig.~\ref{fig:tests-langmuir}, Table~\ref{tab:tests-waves}). At $\nu_e=10\omega_{pe}$, the hydrodynamic wave carries the fraction
$1-e^{-\nu_e\Delta t/2}\approx0.44$ of the electron mass during the field step, which pushes it with the same velocity map as the particles. This case therefore also tests the field response of the
hydrodynamic wave. The frequencies of UGKWP-EC differ from those of DVM-Poisson by at most $0.21\%$ and the damping rates by at most $0.009\omega_{pe}$ (Table~\ref{tab:tests-waves}). At $\nu_e=10\omega_{pe}$, the difference of $A(t)$ of UGKWP-EC from DVM-Poisson is less than
half that of UGKWP-PHM (Fig.~\ref{fig:tests-langmuir}(f)). For $\nu_e=0$ and $\omega_{pe}$, the difference of UGKWP-PHM is mainly an oscillation at a frequency close to
$ck=20$ (Fig.~\ref{fig:tests-langmuir}(d), (e)). It is the oscillation of the cleaning field of the perfectly hyperbolic
Maxwell equations, which travels at the speed of light and is undamped for $\sigma_\phi=0$.

The parallel whistler wave propagates along the uniform magnetic field $\vb B_0=0.25\,\hat{\vb y}$ in a collisionless plasma
with $m_i/m_e=100$ and $T_e=T_i=1.25$, so that $\omega_{pe}/\Omega_{ce}=4$. The periodic domain has the length $2\pi/k$ with
$kd_e=1.25$, where $d_e=c/\omega_{pe}$ is the electron inertial length. The other parameters, the number of cells and the
reference number of particles per cell are those of the electron plasma wave. The computation starts from the cold-plasma
eigenmode with $|\vb B_\perp|=0.01B_0$. The frequency $\omega$ and the damping rate $\gamma$ are obtained by fitting
$A_1e^{-\mathrm i(\omega+\mathrm i\gamma)t}$ to the amplitude of $B_x+\mathrm iB_z$ at the wavenumber $k$ over $10\le t\le80$.
In a cold plasma, the wave would oscillate undamped at $\omega=0.1519$. In the warm plasma, the thermal motion of the
electrons along $\vb B_0$ lowers this frequency, and the electrons in cyclotron resonance with the wave absorb its energy.
With a time step of about 11 light-crossing times of a cell, UGKWP-EC reproduces both effects (Table~\ref{tab:tests-waves}).

\begin{figure}[t]
\centering
\includegraphics[width=\textwidth]{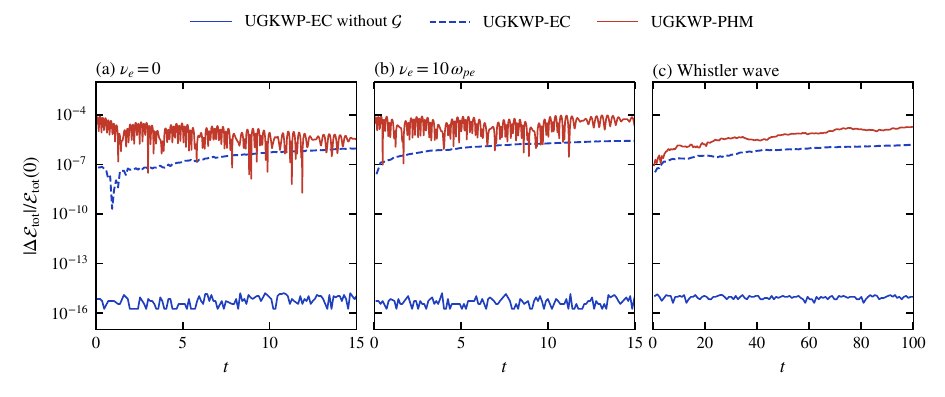}
\caption{Relative change $|\Delta\Ecal_{\rm tot}|/\Ecal_{\rm tot}(0)$ of the total energy, $\Delta\Ecal_{\rm tot}=
\Ecal_{\rm tot}(t)-\Ecal_{\rm tot}(0)$, for (a) the electron plasma wave with $\nu_e=0$, carried by particles, (b) the electron plasma wave with
$\nu_e=10\omega_{pe}$, carried by particles and wave, and (c) the whistler wave, computed by UGKWP-EC without and
with the Gauss correction $\Gop$ and by UGKWP-PHM.}
\label{fig:tests-energy}
\end{figure}

Without the Gauss correction, the computations of the three electron plasma waves and of the whistler wave conserve
the total energy $\Ecal_{\rm tot}$ of Section~\ref{sec:properties} to rounding error, with particles alone and with
particles and wave (Proposition~\ref{prop:energy}(iii), Fig.~\ref{fig:tests-energy}). With the correction, the change of
$\Ecal_{\rm tot}$ is the field-energy change of $\Gop$ (Proposition~\ref{prop:gauss}(iii)) and stays below $3\times10^{-6}$ of
$\Ecal_{\rm tot}$, whereas that of UGKWP-PHM reaches $10^{-4}$.

\begin{figure}[t]
\centering
\includegraphics[width=\textwidth]{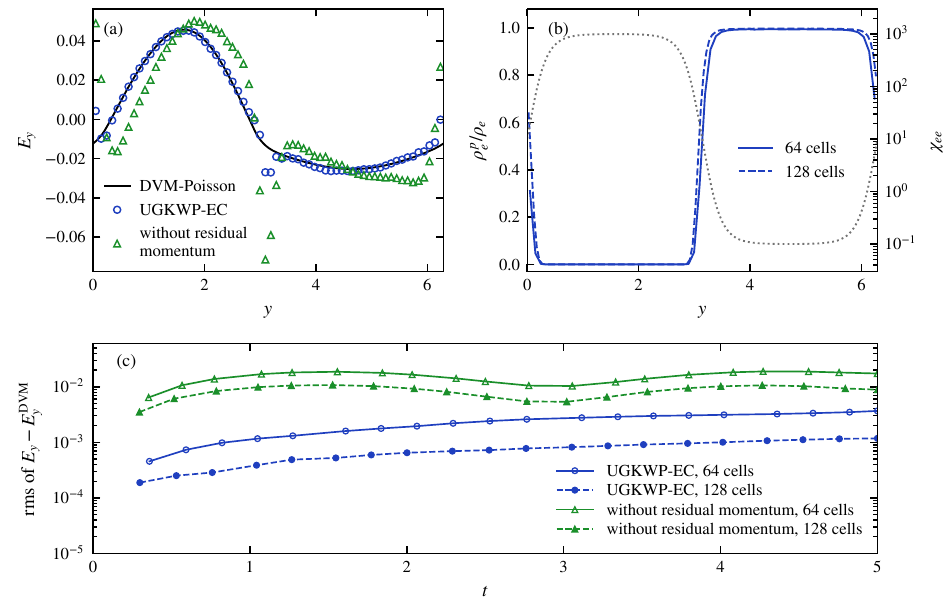}
\caption{Electron plasma wave with the electron collision coefficient \eqref{eq:tests-coex-chi}: (a) $E_y$ at $t=5$ on 64
cells, (b) fraction $\rho^p_e/\rho_e$ of the electron mass carried by particles at $t\approx2.5$ on 64 and 128 cells (left
axis) and $\chi_{ee}(y)$ (dotted, right axis), and (c) rms over the cells of the difference of $E_y$ from DVM-Poisson.
``Without residual momentum'' denotes UGKWP-EC with $\varpi_{e,i}=1$ in \eqref{eq:closure}.}
\label{fig:tests-coex}
\end{figure}

Finally, the electron plasma wave is computed on 64 and 128 cells with the electron collision frequency $\nu_e=\chi_{ee}(y)n_e$,
\begin{equation}
 \chi_{ee}(y)=0.1\times10^{4\sigma(y)},\qquad \sigma(y)=\frac12+\frac{\tanh(3\sin y)}{2\tanh3},
 \label{eq:tests-coex-chi}
\end{equation}
where $\chi_{ee}$ ranges from $0.1$ at $y=3\pi/2$ to $1000$ at $y=\pi/2$, with transitions at $y=0$ and $y=\pi$
(Fig.~\ref{fig:tests-coex}(b)). The reference is DVM-Poisson with the same $\chi_{ee}(y)$ and the cells and velocity points given above.

On 64 cells, $\nu_e\Delta t$ lies between $0.01$ and $120$, so that the particles carry the electron mass where the collisions
are weak and the hydrodynamic wave carries it where they dominate (Fig.~\ref{fig:tests-coex}(b)). The electric field $E_y$ agrees with DVM-Poisson except near the transitions, where the difference decreases as the mesh is refined (Fig.~\ref{fig:tests-coex}(a), (c)).

At the transitions, where the particle fraction varies in space, the analytic fluxes bring momentum into the cell averages but
not into the particles, and this momentum is assigned to the wave through the residual momentum in \eqref{eq:closure}. Without
the residual momentum in the wave, that is, with $\varpi_{e,i}=1$ in \eqref{eq:closure}, $E_y$ departs from DVM-Poisson at both transitions, and on 128 cells this departure remains larger than the
difference of UGKWP-EC on 64 cells (Fig.~\ref{fig:tests-coex}(a), (c)).

\subsection{Multiscale Brio--Wu problem}\label{sec:tests-briowu}
\begin{figure}[p]
\centering
\includegraphics[width=\textwidth]{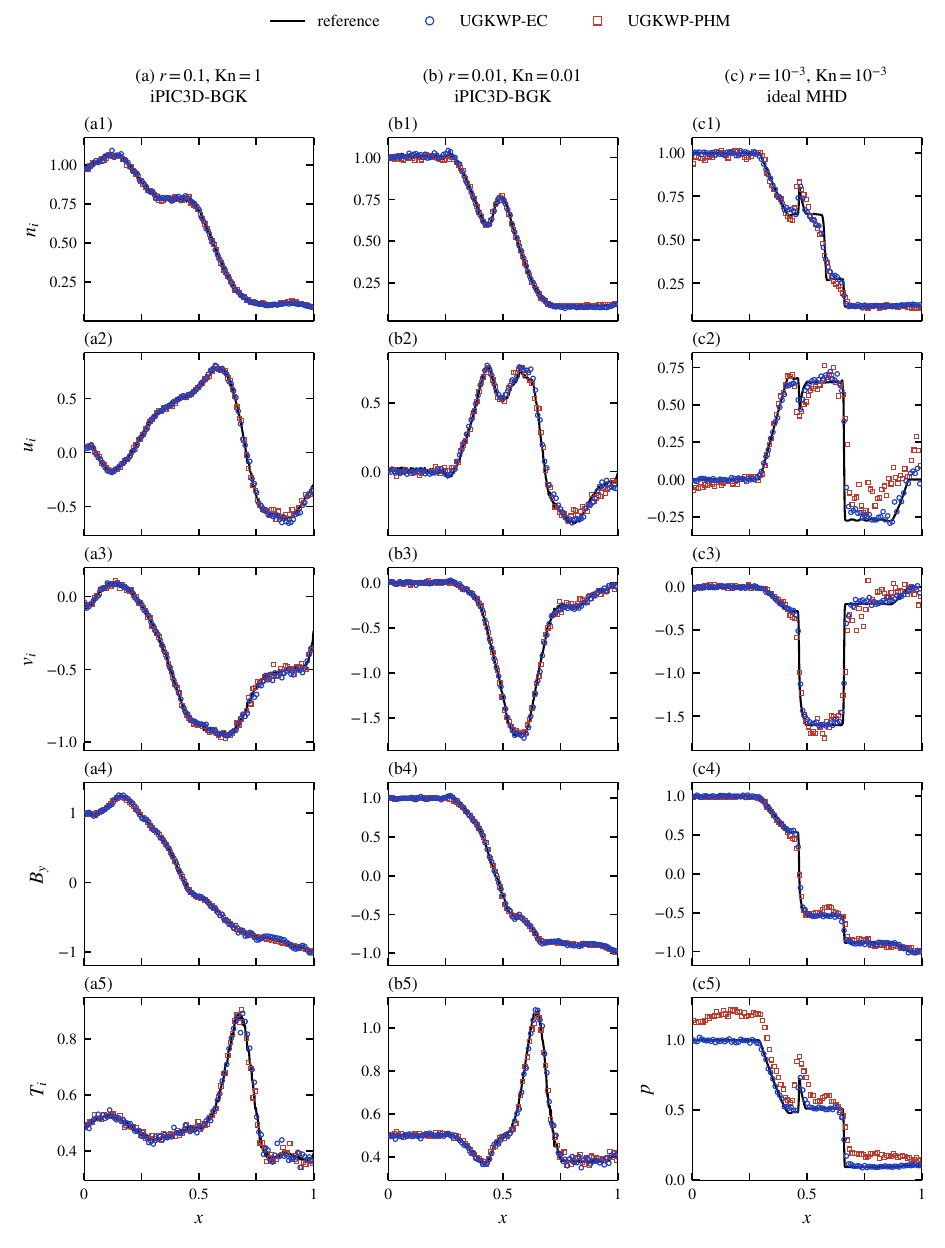}
\caption{Brio--Wu problem at $t=0.12$: (a) $r=0.1$, ${\rm Kn}=1$ and (b) $r=0.01$, ${\rm Kn}=0.01$, with iPIC3D-BGK as the
reference, and (c) $r=10^{-3}$, ${\rm Kn}=10^{-3}$, with ideal MHD (HLLD, 400 cells). Rows 1--5: ion density, ion velocity
components $u_i$ and $v_i$ along $x$ and $y$, $B_y$, and ion temperature $T_i$, except in (c5), which shows the thermal pressure
$p=n_iT_i+n_eT_e$, as ideal MHD has no separate ion temperature. Symbols mark every fourth cell.}
\label{fig:tests-bw}
\end{figure}

The Brio--Wu problem \cite{Brio1988} for electrons and ions \cite{Shumlak2003} tests whether the method follows kinetic solutions at finite Larmor radius and mean free
path and approaches ideal MHD as both decrease. In this subsection, $x$ is the coordinate normal to the discontinuity. Initially, the electrons, with $m_e=0.01$, and the ions,
with $m_i=1$, are Maxwellian with zero mean velocity, and
\[
 (n_s,T_s,\vb B)=\begin{cases}(1,\,0.5,\,(0.75,1,0)), & x<0.5,\\ (0.125,\,0.4,\,(0.75,-1,0)), & x>0.5,\end{cases}\qquad \vb E=\vb 0.
\]
In the normalization of \cite{LiuDUGKS2026}, with the Knudsen number ${\rm Kn}$, the normalized Larmor
radius $r$ and the normalized Debye length $\lambda$,
\[
 \kappa_i=\frac1r,\quad \kappa_e=-\frac{1}{m_er},\quad \epsilon_0=\lambda^2,\quad \mu_0^{-1}=\lambda^2c^2,\quad
 \chi_{ii}=\frac{1}{\rm Kn},\quad \chi_{ee}=\frac{1}{{\rm Kn}\sqrt{m_e}},
\]
with $\lambda=0.01$ and $c=100$. The three settings $(r,{\rm Kn})=(0.1,1)$, $(0.01,0.01)$ and $(10^{-3},10^{-3})$ range from nearly collisionless
ions to the magnetohydrodynamic limit, with $\omega_{pe}\Delta t$ from $2.0$ to $180$, $\Omega_{ce}\Delta t$ from $0.25$ to $23$ and
$c\Delta t/\Delta x$ from $7.2$ to $7.9$. The domain $[0,1]$ has 400 cells, the reference number of particles per cell is 8000 for each species,
and the computations run to $t=0.12$.

The first two settings are compared with kinetic solutions of iPIC3D \cite{Markidis2010} extended by a BGK operator
(iPIC3D-BGK), and the third with ideal MHD. After every push, iPIC3D-BGK replaces each particle by a sample of the
Maxwellian of its species in its cell with the probability $1-e^{-\nu_s\Delta t}$, where $\nu_s=\chi_{ss}n_s$ is the
collision frequency there. A common shift of the particle velocities and a common scaling of their deviations from the mean then conserve the
momentum and the energy of the species in the cell. The iPIC3D-BGK computations use the same domain, a 10 times finer mesh, a 22 times smaller time step and $1024$ particles per species and cell, and their profiles are averaged onto
the 400 cells. The ideal-MHD solution is computed by the HLLD solver \cite{Miyoshi2005} on the same 400 cells, with $\gamma=5/3$ and
the mass density $m_in_i+m_en_e$ and pressure $n_iT_i+n_eT_e$ of the initial states.

As $r$ and Kn decrease from (a) to (c), the smooth profiles of the nearly collisionless ions steepen toward the
discontinuities of ideal MHD (Fig.~\ref{fig:tests-bw}). Like UGKWP-PHM, UGKWP-EC reproduces the profiles of iPIC3D-BGK
for $r=0.1$ and $0.01$, including the peak of the ion temperature (Fig.~\ref{fig:tests-bw}(a), (b)).  For $r=10^{-3}$, it places the fast rarefactions, the
slow compound wave, the contact discontinuity and the slow shock of ideal MHD at the positions of the HLLD solution and
reproduces the peak of the compound wave (Fig.~\ref{fig:tests-bw}(c)). The pressure $p=n_iT_i+n_eT_e$ of UGKWP-EC keeps its initial values where the MHD solution does, whereas that of
UGKWP-PHM exceeds the MHD solution across the domain (Fig.~\ref{fig:tests-bw}(c5)).

In the MHD limit, the relative residual of Gauss's law stays below $10^{-3}$ with the Gauss correction and reaches $10^{-2}$
without it.

\FloatBarrier
\subsection{Orszag--Tang vortex}\label{sec:tests-ot}
The Orszag--Tang vortex \cite{Orszag1979} tests the method in two dimensions, on a flow that forms shocks and thin current sheets, and
shows how the number of particles and the computational cost change with the Knudsen number. It is computed on the periodic domain $[0,2\pi]^2$ in
the normalization of Section~\ref{sec:tests-briowu} with $m_i=1$, $m_e=0.04$, $\lambda=0.05$, $c=20$ and $r=0.2$, so
that the ion inertial length is $0.12$. Initially,
\[
 \rho=\gamma^2,\quad p=\gamma,\quad \vb U=(-\sin y,\sin x,0),\quad \vb B=(-\sin y,\sin2x,0),\quad
 \vb E=-\vb U\times\vb B,
\]
where $\gamma=5/3$, $\rho=\rho_i+\rho_e$ is the mass density, $p=n_iT_i+n_eT_e$ the pressure and $\vb U$ the velocity of
the plasma, with $n_i=n_e$ and $T_i=T_e$. Both species move with the in-plane velocity of $\vb U$. Their out-of-plane velocities give the plasma zero out-of-plane
momentum and carry the current,
\[
 \rho_iU_{i,z}+\rho_eU_{e,z}=0,\qquad \kappa_i\rho_iU_{i,z}+\kappa_e\rho_eU_{e,z}=\frac{(\nabla\times\vb B)_z}{\mu_0}.
\]
The species have the collision coefficients $\chi_{ii}=1/{\rm Kn}$ and $\chi_{ee}=1/({\rm Kn}\sqrt{m_e})$, for ${\rm Kn}=10^{-3}$, $10^{-2}$ and $10^{-1}$. The mesh has $256^2$ square cells of side $h$, a fifth of the ion inertial length. The reference number of particles per cell
is 256 for each species, and the computations run to $t=3$.
The flow structures at ${\rm Kn}=10^{-3}$ and $10^{-1}$ are compared with \mbox{UGKWP-PHM} and with a reference for each
Knudsen number, all computed on the same $256^2$ mesh.
For ${\rm Kn}=10^{-3}$, where the mean free paths are about $h/100$, the reference is the five-moment two-fluid solution
computed by the Gkeyll code \cite{Gkeyll} with the finite-volume wave-propagation
scheme of \cite{Hakim2006} and the locally implicit source update of \cite{Wang2020}. For
${\rm Kn}=10^{-1}$, the reference is iPIC3D-BGK (Section~\ref{sec:tests-briowu}) with 4096 particles per species and cell and a fixed time step of about
half the mean step of UGKWP-EC. The implicit scheme of iPIC3D-BGK solves the field
equations for the electric field at $t^n+\theta\Delta t$, with $\theta=1/2$.

\begin{figure}[tbp]
\centering
\includegraphics[width=0.8\textwidth]{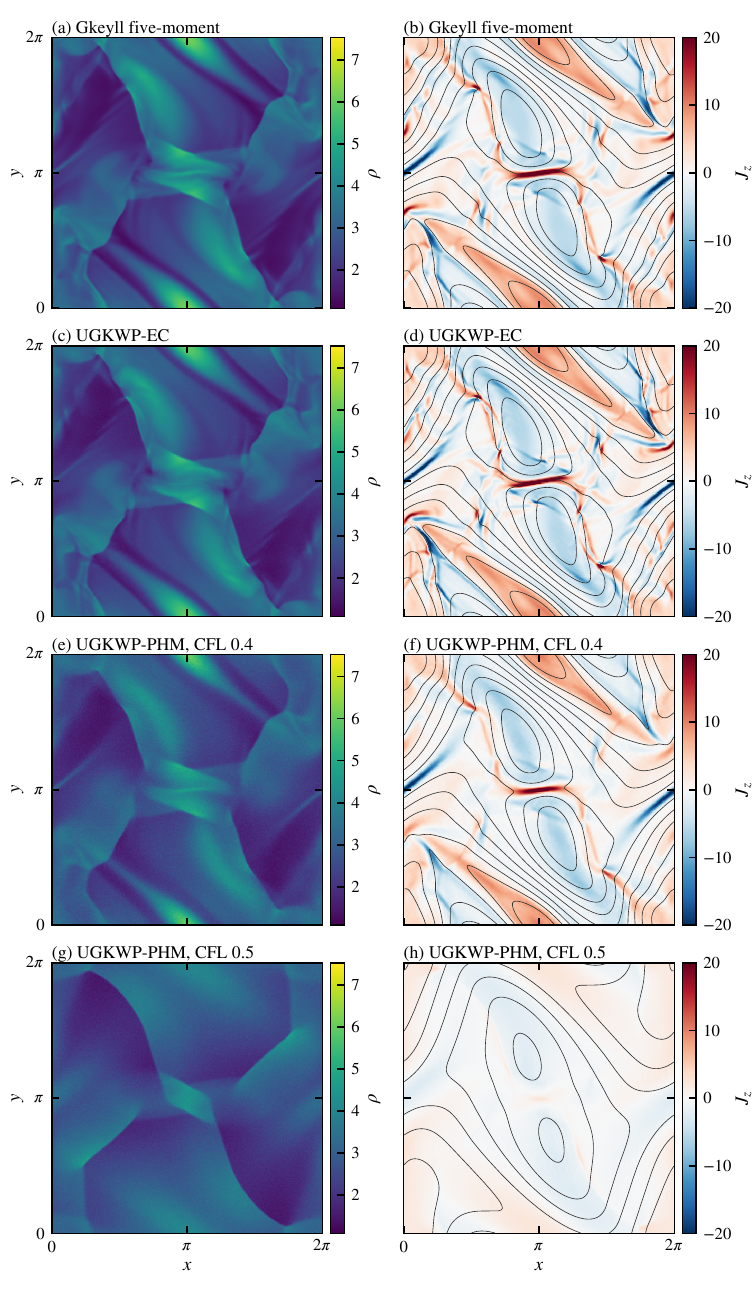}
\caption{Orszag--Tang vortex, ${\rm Kn}=10^{-3}$, $256^2$ cells, $t=3$: (a), (c), (e), (g) total mass density and (b),
(d), (f), (h) $J_z=(\nabla\times\vb B)_z/\mu_0$ with in-plane field lines (contours of $A_z$), for (a), (b) the
Gkeyll five-moment reference, (c), (d) UGKWP-EC, (e), (f) UGKWP-PHM at CFL $0.4$ and (g), (h) UGKWP-PHM at CFL $0.5$.
The color scale of $J_z$ is saturated at $|J_z|=20$.}
\label{fig:tests-ot-maps-kn1e-3}
\end{figure}

\begin{figure}[tbp]
\centering
\includegraphics[width=0.8\textwidth]{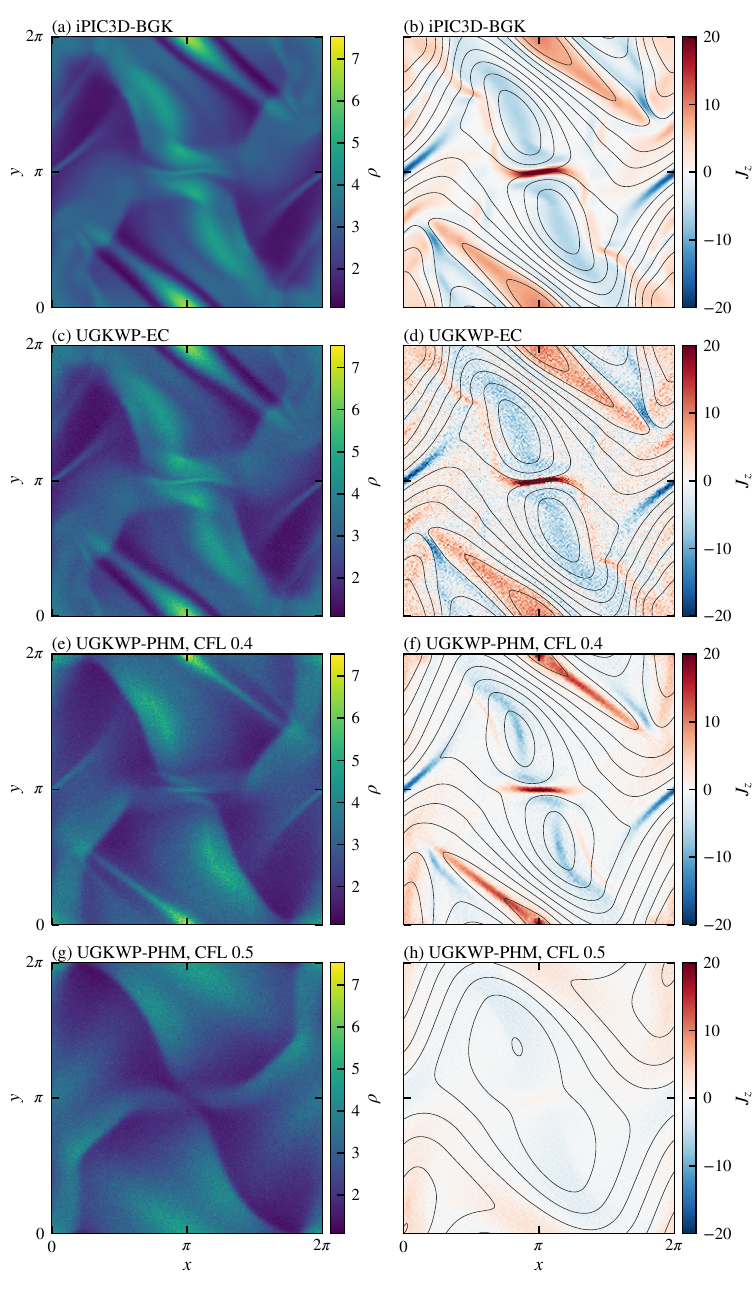}
\caption{Orszag--Tang vortex, ${\rm Kn}=10^{-1}$, $256^2$ cells, $t=3$: (a), (c), (e), (g) total mass density and (b),
(d), (f), (h) $J_z=(\nabla\times\vb B)_z/\mu_0$ with in-plane field lines (contours of $A_z$), for (a), (b) the
iPIC3D-BGK reference, (c), (d) UGKWP-EC, (e), (f) UGKWP-PHM at CFL $0.4$ and (g), (h) UGKWP-PHM at CFL $0.5$. The color
scale of $J_z$ is saturated at $|J_z|=20$.}
\label{fig:tests-ot-maps-kn1e-1}
\end{figure}

\begin{figure}[tbp]
\centering
\includegraphics[width=\textwidth]{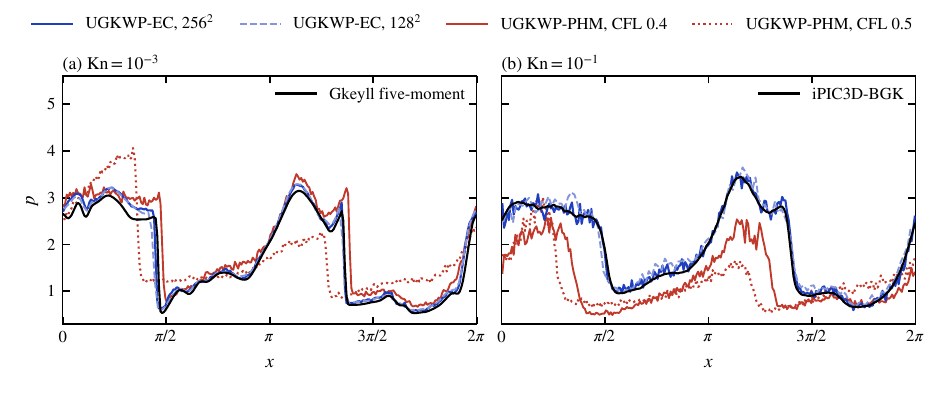}
\caption{Orszag--Tang vortex, $t=3$: pressure $p$ along $y=0.625\pi$ for (a) ${\rm Kn}=10^{-3}$ and (b)
${\rm Kn}=10^{-1}$, with the Gkeyll five-moment solution and iPIC3D-BGK as the references.}
\label{fig:tests-ot-cut}
\end{figure}

For both Knudsen numbers, UGKWP-EC forms the shocks, the central current sheet and the magnetic islands of the reference
(Figs.~\ref{fig:tests-ot-maps-kn1e-3} and \ref{fig:tests-ot-maps-kn1e-1}). At ${\rm Kn}=10^{-1}$, where the mean free paths
of the ions and the electrons are about $h$, the shocks are wider than at ${\rm Kn}=10^{-3}$ and the thin current sheets
along them are much weaker. At both Knudsen numbers, the pressure of UGKWP-EC along $y=0.625\pi$ changes little when the mesh is refined from $128^2$ to $256^2$ cells (Fig.~\ref{fig:tests-ot-cut}).

The total energy of UGKWP-EC changes only by the field-energy change of $\Gop$ (Proposition~\ref{prop:gauss}(iii)), by at
most $0.32\%$ of its initial value, whereas that of UGKWP-PHM changes by $3.8\%$ to $35\%$ at the two CFL numbers. At CFL $0.5$,
the dissipation of the Maxwell solver of UGKWP-PHM reduces the magnetic energy to less than half of its initial value by $t=3$ and smears out
the shocks and the current sheets (Figs.~\ref{fig:tests-ot-maps-kn1e-3}(g), (h) and \ref{fig:tests-ot-maps-kn1e-1}(g), (h)). At
${\rm Kn}=10^{-1}$, the electrons of UGKWP-PHM lose more than half of their thermal energy by $t=3$, so that its pressure is lower
at both CFL numbers and its shocks are displaced (Fig.~\ref{fig:tests-ot-cut}(b)). In the solution of this problem, the total thermal energy of each species does not fall below its initial value, because the
entropy of each species does not decrease and the initial Maxwellian of uniform density and temperature has the largest
entropy for its total mass and thermal energy. At $t=3$, the thermal energy of each species lies above its initial value in UGKWP-EC and in the references.

At ${\rm Kn}=10^{-3}$, the wave carries
almost all of the electrons, and ion particles remain only in the low-density regions. The mean numbers of particles per cell over the run are 2 for
the electrons and 8 for the ions, compared with the reference number of 256. At ${\rm Kn}=10^{-1}$, the particles carry
most of the mass of both species, and their numbers are close to the reference number.

The CPU time of UGKWP-EC is compared at the three Knudsen numbers with that of iPIC3D-BGK, which represents both species by
particles at every Knudsen number. The iPIC3D-BGK computations use the same mesh, $256$ particles per species and cell and
$\theta=1$, because with $\theta=1/2$ they are unstable at $\omega_{pe}\Delta t\approx2.5$. They solve the field equations by GMRES to the same
relative residual $10^{-8}$ and take the time step from \eqref{eq:tests-cfl} with the same CFL number. Although
iPIC3D-BGK takes fewer time steps than UGKWP-EC, it needs $1.4$ to $4.2$ times as much CPU time, and the ratio grows
as the Knudsen number decreases and the wave replaces the particles (Table~\ref{tab:tests-ot-cost}).
At CFL $0.4$, UGKWP-PHM needs $4.0$ to $5.7$ times the CPU time of UGKWP-EC.

\begin{table}[ht]
\centering
\caption{Orszag--Tang vortex on $256^2$ cells: number of steps, CPU time in core-hours and ratio of the CPU time to that
of UGKWP-EC.}
\label{tab:tests-ot-cost}
\footnotesize
\begin{tabular*}{\textwidth}{@{\extracolsep{\fill}}c cd{1.1} cd{2.1}d{1.1} cd{2.1}d{1.1}@{}}
\toprule
 & \multicolumn{2}{c}{UGKWP-EC} & \multicolumn{3}{c}{UGKWP-PHM, CFL $0.4$} & \multicolumn{3}{c}{iPIC3D-BGK}\\
\cmidrule(lr){2-3}\cmidrule(lr){4-6}\cmidrule(l){7-9}
${\rm Kn}$ & Steps & \colhead{CPU time} & Steps & \colhead{CPU time} & \colhead{Ratio} & Steps & \colhead{CPU time}
 & \colhead{Ratio}\\
\midrule
$10^{-3}$ & $977$ & 2.6 & $6112$ & 10.3 & 4.0 & $872$ & 10.8 & 4.2\\
$10^{-2}$ & $954$ & 4.4 & $6112$ & 25.2 & 5.7 & $875$ & 10.3 & 2.3\\
$10^{-1}$ & $952$ & 6.4 & $6112$ & 30.9 & 4.8 & $849$ & 8.9 & 1.4\\
\bottomrule
\end{tabular*}
\end{table}

\FloatBarrier
\subsection{Magnetic reconnection}\label{sec:tests-gem}
\begin{figure}[t]
\centering
\includegraphics[width=\textwidth]{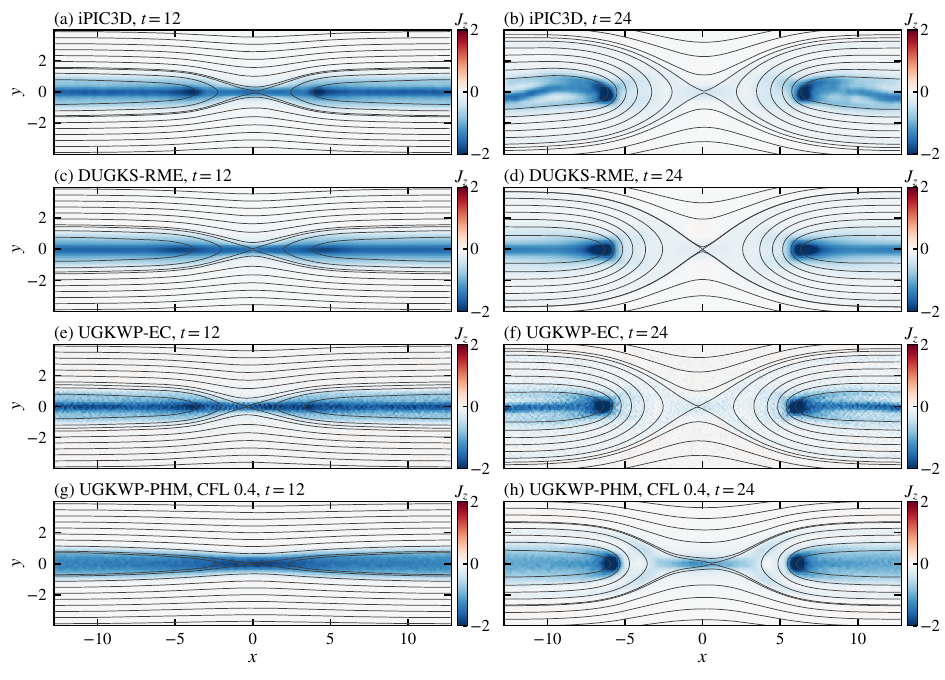}
\caption{GEM problem, $c/v_A=20$, $256\times128$ cells: out-of-plane current $J_z=(\nabla\times\vb B)_z$, computed from the
magnetic field of every code by the same central differences, and in-plane magnetic field lines (contours of $A_z$) for
$|y|\le4$ at $t=12$ and $24$. The color scale is saturated at $|J_z|=2$.}
\label{fig:tests-gem-jz}
\end{figure}

The collisionless Geospace Environmental Modeling (GEM) reconnection problem \cite{Birn2001} tests the method on Hall
reconnection, with time steps longer than the
light-crossing time of a cell. It is computed on the domain $[-12.8,12.8]\times[-6.4,6.4]$ with $m_i/m_e=25$ and $c/v_A=20$ or $10$, where
$v_A$ is the Alfv\'en speed. In units of the ion inertial length $d_i$, the inverse ion cyclotron frequency
$\Omega_{ci}^{-1}$, $B_0$ and $n_0$, in which $v_A=1$, the coefficients are $\kappa_i=1$, $\kappa_e=-25$, $\mu_0=1$ and
$\epsilon_0=(v_A/c)^2$. Each species consists of a
Harris sheet population and a background population, both Maxwellian with $T_i=5/12$ and
$T_e=1/12$, with the densities and velocities
\[
 n^{\rm sh}_s=\mathrm{sech}^2(2y),\quad \vb U^{\rm sh}_s=U_s\hat{\vb z},\qquad n^{\rm bg}_s=0.2,\quad \vb U^{\rm bg}_s=\vb 0,
\]
where $U_i=-5/3$ and $U_e=1/3$. Initially, $\vb E=\vb 0$ and $\vb B=\nabla\times(A_z\hat{\vb z})$ with
\[
 A_z=0.5\ln\cosh(2y)+0.1\cos\frac{\pi x}{12.8}\cos\frac{\pi y}{12.8}+0.08\,e^{-(x^2+y^2)/4},
\]
the sum of the Harris field, the GEM perturbation and a Gaussian perturbation at the X point. The particles carry the
current of the Harris field only. The domain is periodic in $x$ and bounded at $y=\pm6.4$ by perfectly conducting walls
that reflect the particles specularly. Both species are collisionless and therefore carried by particles only, with 256
particles per cell for each species and population, and the mesh has $256\times128$ square cells of side $h$.
The references, computed for both values of $c/v_A$, are iPIC3D on $256\times128$ cells with the same initial particle
populations and perturbation, and DUGKS-RME \cite{LiuDUGKS2026} on $256\times128$ cells with $40^3$ discrete velocity points per species and the
conservative semi-Lagrangian update of \cite{Liu2025} in velocity space. The reconnected flux $\Psi=A_z(\mathrm X)-A_z(\mathrm O)$ is the magnetic flux
between the X point at the center of the domain and the O point at its periodic boundary, with the flux function $A_z$
on the midplane.

\begin{figure}[t]
\centering
\includegraphics[width=\textwidth]{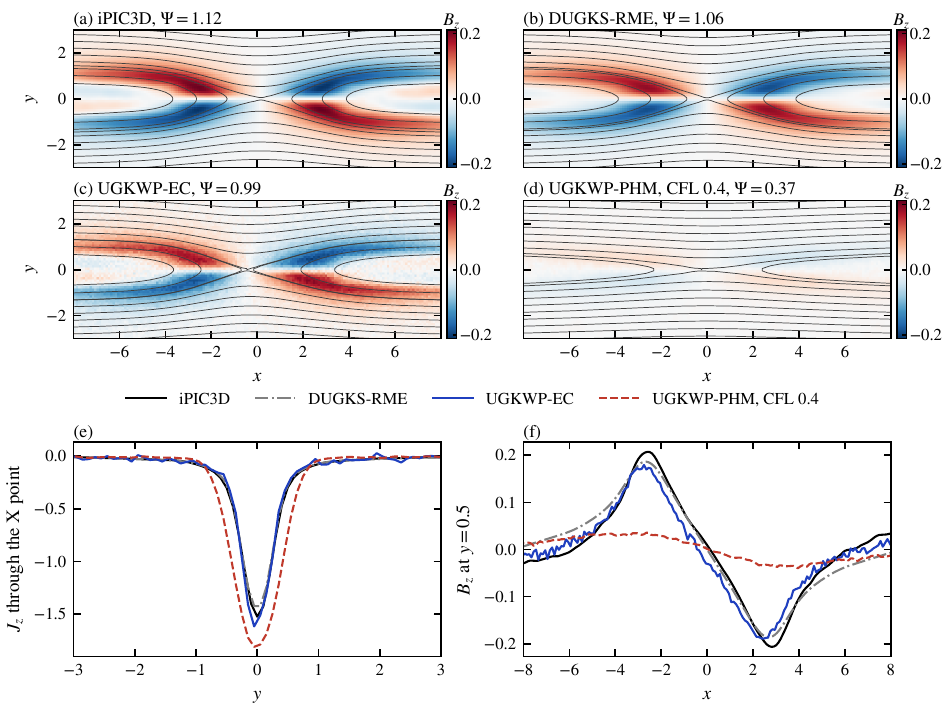}
\caption{GEM problem, $c/v_A=20$, $256\times128$ cells, $t=12$: (a)--(d) out-of-plane magnetic field $B_z$ and in-plane
field lines around the X point, with the reconnected flux $\Psi$, (e) $J_z$ along $y$ through the
X point, averaged over $|x|\le0.5$, and (f) $B_z$ along $y=0.5$.}
\label{fig:tests-gem-bz}
\end{figure}

\begin{figure}[t]
\centering
\includegraphics[width=\textwidth]{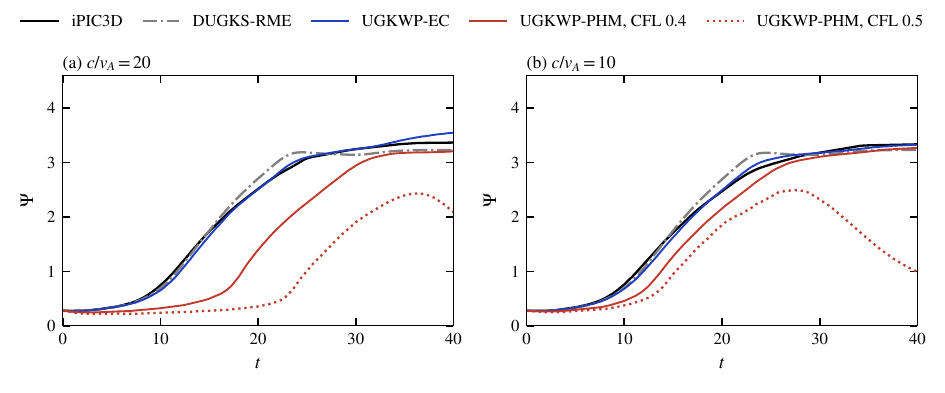}
\caption{GEM problem: reconnected flux $\Psi(t)$ for (a) $c/v_A=20$ and (b) $c/v_A=10$, on $256\times128$ cells.}
\label{fig:tests-gem-psi}
\end{figure}

\begin{figure}[t]
\centering
\includegraphics[width=\textwidth]{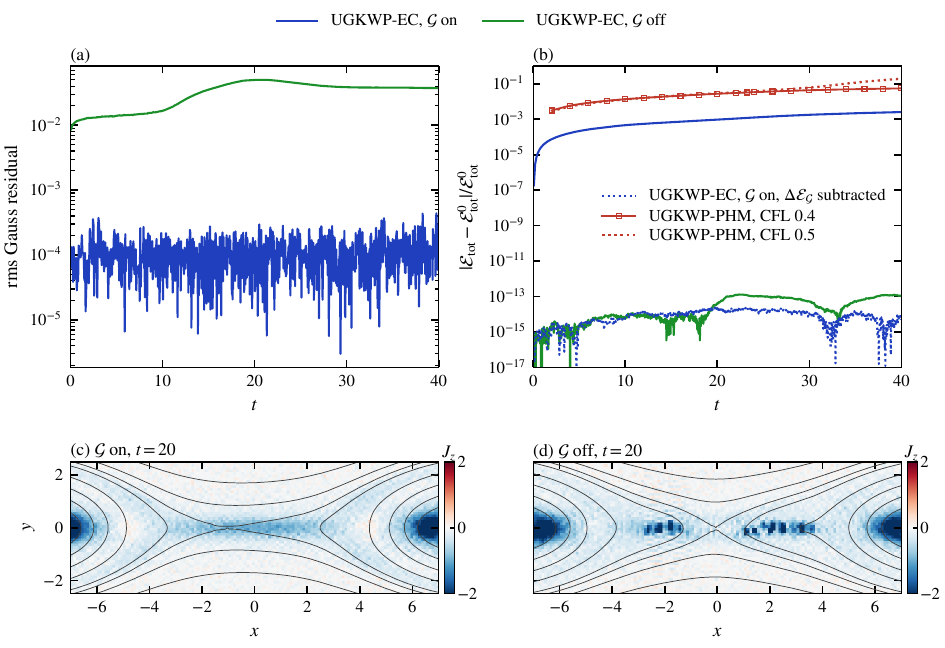}
\caption{GEM problem, $c/v_A=20$, $256\times128$ cells, with and without the Gauss correction. (a) The rms over the interior nodes of the residual of the weak form of Gauss's law, divided by the nodal area, in units of
$\kappa_in_0$. (b) Relative change of the discrete total energy of
UGKWP-EC without and with $\Gop$, the latter also after subtracting the accumulated field-energy change
$\Delta\Ecal_{\Gop}$ of $\Gop$ (blue dotted), and of the total energy of UGKWP-PHM at CFL $0.4$ and at CFL $0.5$ (red
dotted). (c), (d) $J_z$ and
in-plane field lines at $t=20$ with and without $\Gop$, with the color scale of Fig.~\ref{fig:tests-gem-jz}.}
\label{fig:tests-gem-g}
\end{figure}

At $t=12$ and $24$, when the reconnected flux of the references is about $1$ and $3$, UGKWP-EC reproduces the
thinning of the current sheet at the X point, the opening of the separatrices and the islands of iPIC3D and DUGKS-RME in shape and position (Fig.~\ref{fig:tests-gem-jz}). The
out-of-plane magnetic field of UGKWP-EC forms the Hall quadrupole of collisionless reconnection, with extremes along $y=0.5$ close to those of DUGKS-RME (Fig.~\ref{fig:tests-gem-bz}(a)--(d), (f)). Through the X point, the current
layer of UGKWP-EC is about as thin as in the references (Fig.~\ref{fig:tests-gem-bz}(e)). Reconnection starts later in
UGKWP-PHM at CFL $0.4$, and more so at $c/v_A=20$ than at $10$ (Table~\ref{tab:tests-gem}), because the dissipation of its
explicit Maxwell solver increases with $c$ and diffuses the initial current sheet. At $t=12$, its current layer through the X
point is still about as wide as the initial Harris sheet, its reconnected flux is about a third of that of the references, and
its Hall field is still weak (Fig.~\ref{fig:tests-gem-bz}).

\begin{table}[t]
\centering
\caption{GEM problem on $256\times128$ cells: time at which $\Psi=1$, maximum of $\mathrm d\Psi/\mathrm dt$ averaged over intervals of length $1$, $\Psi(40)$, and relative change of
the total energy at $t=40$.}
\label{tab:tests-gem}
\footnotesize
\setlength{\tabcolsep}{3pt}
\begin{tabular*}{\textwidth}{@{\extracolsep{\fill}}l d{2.1}d{1.3}d{1.2}d{3.2} d{2.1}d{1.3}d{1.2}d{3.2}@{}}
\toprule
 & \multicolumn{4}{c}{$c/v_A=20$} & \multicolumn{4}{c}{$c/v_A=10$}\\
\cmidrule(lr){2-5}\cmidrule(l){6-9}
Method & \colhead{$t(\Psi=1)$} & \colhead{$\max\dot\Psi$} & \colhead{$\Psi(40)$} & \colhead{$\Delta\Ecal_{\rm tot}$ (\%)}
 & \colhead{$t(\Psi=1)$} & \colhead{$\max\dot\Psi$} & \colhead{$\Psi(40)$} & \colhead{$\Delta\Ecal_{\rm tot}$ (\%)}\\
\midrule
iPIC3D & 11.4 & 0.213 & 3.37 & -0.47 & 11.3 & 0.200 & 3.34 & -0.67\\
DUGKS-RME & 11.7 & 0.239 & 3.23 & +0.18 & 11.7 & 0.235 & 3.24 & +0.16\\
UGKWP-EC & 12.0 & 0.228 & 3.55 & -0.25 & 12.0 & 0.213 & 3.33 & -0.27\\
UGKWP-PHM, CFL $0.4$ & 18.4 & 0.270 & 3.21 & -5.47 & 13.8 & 0.228 & 3.27 & -4.28\\
UGKWP-PHM, CFL $0.5$ & 25.0 & 0.244 & 2.09 & -19.12 & 15.2 & 0.221 & 0.99 & -23.66\\
\bottomrule
\end{tabular*}
\end{table}

For both values of
$c/v_A$, the reconnected flux of UGKWP-EC stays close to those of the references, and its peak rate lies between theirs
(Fig.~\ref{fig:tests-gem-psi}, Table~\ref{tab:tests-gem}). UGKWP-EC takes about 2200 steps for both values of $c/v_A$, with time steps of about 3.7 and 1.8 light-crossing
times of a cell. At CFL $0.4$, UGKWP-PHM needs $11.5$ and $6.3$ times the CPU time of UGKWP-EC, and DUGKS-RME needs $132$ and $137$ times (Table~\ref{tab:tests-gem-cost}). UGKWP-PHM loses about $5\%$ of its total energy, and at CFL $0.5$ it reconnects still later and loses more
than $19\%$, whereas the total energy of UGKWP-EC changes by less than $0.3\%$ (Table~\ref{tab:tests-gem}).

\begin{table}[t]
\centering
\caption{GEM problem on $256\times128$ cells: number of steps, CPU time in core-hours and ratio of the CPU time to that
of UGKWP-EC.}
\label{tab:tests-gem-cost}
\footnotesize
\begin{tabular*}{\textwidth}{@{\extracolsep{\fill}}c cd{2.1} cd{3.1}d{2.1} ccc@{}}
\toprule
 & \multicolumn{2}{c}{UGKWP-EC} & \multicolumn{3}{c}{UGKWP-PHM, CFL $0.4$} & \multicolumn{3}{c}{DUGKS-RME}\\
\cmidrule(lr){2-3}\cmidrule(lr){4-6}\cmidrule(l){7-9}
$c/v_A$ & Steps & \colhead{CPU time} & Steps & \colhead{CPU time} & \colhead{Ratio} & Steps & CPU time & Ratio\\
\midrule
$20$ & $2183$ & 11.0 & $20000$ & 126.4 & 11.5 & $5142$ & $1452$ & $132$\\
$10$ & $2166$ & 10.6 & $10000$ & 67.1 & 6.3 & $5142$ & $1444$ & $137$\\
\bottomrule
\end{tabular*}
\end{table}

With $\Gop$, the rms of the residual of the weak form of Gauss's law over the interior nodes is two orders of magnitude
below that of the otherwise identical run without $\Gop$ (Fig.~\ref{fig:tests-gem-g}(a)), and the total energy changes only
by the field-energy change of $\Gop$ (Proposition~\ref{prop:gauss}(iii)). Without $\Gop$, the total energy is conserved
(Proposition~\ref{prop:energy}(iii), Fig.~\ref{fig:tests-gem-g}(b)), but the current layer through the X point breaks by
$t=20$ into a row of small current patches a few cells apart, which do not form with $\Gop$ (Fig.~\ref{fig:tests-gem-g}(c),
(d)).
\FloatBarrier
\section{Conclusions}\label{sec:conclusions}

For the BGK--Vlasov--Maxwell system, this paper has presented UGKWP-EC, a semi-implicit UGKWP method in which the particles and the
hydrodynamic wave share one electromagnetic response. The current response of the reformulated Maxwell equation is built from the
implicit-midpoint velocity map that then pushes the particles and maps the wave. This equation is a single linear system
for the electric field, with a unique solution for any time step. As Maxwell's equations are advanced by the implicit midpoint rule with the same current, the field step conserves the
discrete total energy exactly.
The Gauss correction adjusts the electric field and the joint charge of particles and wave while conserving the mass, momentum and energy of every
charged species. The transport steps sample and transport the field-pushed wave and the corrected wave without rebuilding them from
their cell averages.

The field step remains stable without resolving the cyclotron period, the plasma period or the light-crossing time,
so that the
time step can be taken from a CFL condition on the flow and sound speeds of the species. The method reproduces the whistler wave of a warm plasma and the electron plasma wave from Landau damping to collision-dominated oscillations. For an electron plasma wave whose collision frequency varies over
four orders of magnitude in space, it approaches the kinetic reference with mesh refinement. For the drifting plasma of \ref{app:drift}, its rotation field keeps the electrons as close to equilibrium as in the plasma without drift, whereas the rotation with the magnetic field of $t^n$ heats them at $\omega_{pe}\Delta t\approx2.5$. It agrees with particle-in-cell, two-fluid and magnetohydrodynamic references in the multiscale Brio--Wu
problem across the Knudsen number from $1$ to $10^{-3}$ and the normalized Larmor radius from $0.1$ to $10^{-3}$, in the Orszag--Tang
vortex at two Knudsen numbers, where particles and wave coexist in two dimensions, and in collisionless magnetic
reconnection. The Gauss correction reduces the residual of Gauss's law by one to two orders of magnitude, and in the GEM problem it
suppresses the small current patches that form along the current layer without it. In the two-dimensional computations, the UGKWP method with an explicit solver for the perfectly hyperbolic Maxwell
equations needs $4.0$ to $11.5$ times the CPU time of \mbox{UGKWP-EC}, whose time step does not depend on $c$. In the
Orszag--Tang vortex, with the same mesh, reference number of particles per cell, collision operator and CFL condition, iPIC3D-BGK needs $1.4$ to $4.2$ times the
CPU time of UGKWP-EC from ${\rm Kn}=10^{-1}$ to $10^{-3}$, and the saving grows as the wave replaces the particles.
In collisionless magnetic reconnection, DUGKS-RME with $40^3$ discrete velocity points per species needs more than $130$ times the CPU
time of UGKWP-EC.

\section*{Declaration of competing interest}
The authors declare that they have no known competing financial interests or personal relationships that could have
appeared to influence the work reported in this paper.

\section*{Acknowledgements}
The current research is supported by the National Key R\&D Program of China (Grant No.~2022YFA1004500), the National
Natural Science Foundation of China (92371107), and the Hong Kong Research Grants Council (16208324).
\appendix
\setcounter{figure}{0}
\section{Rotation field in a drifting plasma}\label{app:drift}
The field step conserves the total energy for any rotation field (Proposition~\ref{prop:energy}). For a drifting plasma, this
does not bound the perturbations, because the drift can supply them with energy. This appendix first compares computations of a drifting plasma with the rotation field \eqref{eq:frozen-b} and with the magnetic
field $\vb B^n_h=\ms N_B\ms b^n$ of $t^n$, which the particle update of ECsim \cite{Lapenta2017} uses. It then shows that, for a drifting cold plasma, the
time step with \eqref{eq:frozen-b}, without $\ms\Phi$ and $\ms S$, conserves a quadratic form that bounds the perturbations as long as it is
positive definite. The condition for this is the bound $\beta_i<4$ of Section~\ref{sec:rotation}. With the field of $t^n$, the step
loses this invariant.

The plasma is the exact uniform equilibrium of
collisionless electrons and ions with the density $n_0$, the common temperature $T$ and the common drift
\[
 \vb U_e=\vb U_i=\vb V=(V,0,0),\qquad \vb B=(0,0,1),\qquad \vb E=-\vb V\times\vb B,
\]
with $m_i/m_e=25$ and $c^2=400$ in the units of Section~\ref{sec:tests-gem}. The periodic domain of length $L=8$ has 96 cells with 256 particles per species and cell. The computations run to $t=40$
with $T=0.1$, $V=0.5$ and the prescribed time steps
$\Delta t=0.025$ and $0.0125$. At $\Delta t=0.025$, the same plasma without drift, $V=0$, is also computed. The thermal energy of
the electrons is constant in the exact solution, so that its change measures the numerical heating.

\begin{figure}[t]
\centering
\includegraphics[width=\textwidth]{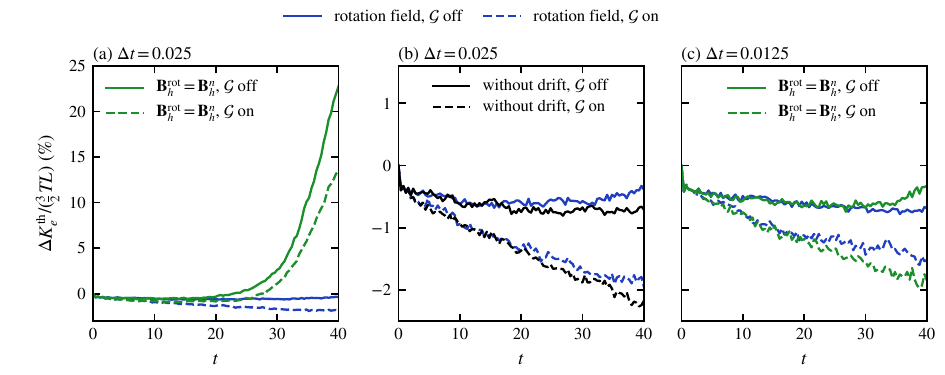}
\caption{Drifting plasma, $T=0.1$, $V=0.5$, 256 particles per species and cell: change of the electron thermal energy in percent of
$\frac32TL$, each without and with the Gauss correction $\Gop$, (a) at $\Delta t=0.025$ with the rotation field
\eqref{eq:frozen-b} and with $\vb B^{\rm rot}_h=\vb B^n_h$, (b) at $\Delta t=0.025$ on a finer scale with the rotation field and
for the plasma without drift ($V=0$), and (c) at $\Delta t=0.0125$ with the rotation field and with $\vb B^{\rm rot}_h=\vb B^n_h$.}
\label{fig:tests-drift}
\end{figure}

At $\Delta t=0.025$, the electrons of the drifting plasma are heated by up to $23\%$ with $\vb B^{\rm rot}_h=\vb B^n_h$, although the
step without the Gauss correction conserves the total energy (Fig.~\ref{fig:tests-drift}(a)). With the rotation field
\eqref{eq:frozen-b}, their thermal energy decreases by at most $2\%$ with and without the Gauss correction and follows that
of the plasma without drift (Fig.~\ref{fig:tests-drift}(b)), so that the drift does not heat them. At $\Delta t=0.0125$, both
choices keep the change of the thermal energy within $2\%$ (Fig.~\ref{fig:tests-drift}(c)).

Consider a uniform, neutral and cold plasma whose species, with the mass densities $\rho_s$ and the charge-to-mass
ratios $\kappa_s$, drift with the velocity $V\hat{\vb x}$ in the magnetic field $B_0\hat{\vb z}$ and the electric field
$VB_0\hat{\vb y}$. A perturbation of this state proportional to $e^{\mathrm iky}$ consists of the displacements $\xi_s$ of the particles along
$\hat{\vb y}$, their velocities $u_s\hat{\vb x}+w_s\hat{\vb y}$, the electric field $E_x\hat{\vb x}+E_y\hat{\vb y}$ and the magnetic field
$b\hat{\vb z}$.
The analysis omits collisions, the Gauss correction and the matrices $\ms\Phi$ and $\ms S$ in \eqref{eq:frozen-b}, and it
uses exact spatial derivatives. Linearizing one time step of UGKWP-EC about the drifting state gives
\begin{gather*}
 \xi^{n+1/2}_s=\xi^n_s+\delta w^n_s,\qquad
 \xi^{n+1}_s=\xi^{n+1/2}_s+\delta w^{n+1}_s,\\
 \bar u_s-u^n_s=\delta\kappa_s\bigl(\bar E_x+B_0\bar w_s\bigr),\qquad
 \bar w_s-w^n_s=\delta\kappa_s\bigl(\bar E_y-B_0\bar u_s-Vb^{\rm rot}\bigr),\\
 \epsilon_0\bigl(\bar E_x-E^n_x\bigr)=\delta\Bigl[\frac{\mathrm ik\bar b}{\mu_0}
 -\sum_s\kappa_s\rho_s\bigl(\bar u_s-\mathrm ikV\xi^{n+1/2}_s\bigr)\Bigr],\\
 \epsilon_0\bigl(\bar E_y-E^n_y\bigr)=-\delta\bar J_y,\qquad
 \bar b=b^n+\delta \mathrm ik\bar E_x,
\end{gather*}
where $b^{\rm rot}=b^n+\delta \mathrm ikE^n_x$, $J_y=\sum_s\kappa_s\rho_sw_s$ and $\bar a=\frac12(a^n+a^{n+1})$ for every
unknown $a$. The step preserves Gauss's law $\epsilon_0E_y=-\sum_s\kappa_s\rho_s\xi_s$, and for perturbations that
satisfy it, $\sum_s\kappa_s\rho_s\xi^{n+1/2}_s=-\epsilon_0E^n_y+\delta J^n_y$. With this relation and the identities
$b^{\rm rot}-\bar b=-\frac12\delta \mathrm ik(E^{n+1}_x-E^n_x)$ and $E^n_y-\bar E_y=\delta\bar J_y/\epsilon_0$, the step
takes the form
\[
 \ms T(\ms y^{n+1}-\ms y^n)=\frac{\Delta t}{2}\,\ms L(\ms y^{n+1}+\ms y^n),\qquad \ms y=(u_s,w_s,E_x,E_y,b),
\]
where $\ms L$ is the matrix of the linearized cold-fluid equations
\[
 \begin{gathered}
 \frac{\dd u_s}{\dd t}=\kappa_s(E_x+B_0w_s),\qquad
 \frac{\dd w_s}{\dd t}=\kappa_s(E_y-B_0u_s-Vb),\qquad
 \frac{\dd b}{\dd t}=\mathrm ikE_x,\\
 \epsilon_0\frac{\dd E_x}{\dd t}=\frac{\mathrm ikb}{\mu_0}-\sum_s\kappa_s\rho_su_s-\mathrm ikV\epsilon_0E_y,\qquad
 \epsilon_0\frac{\dd E_y}{\dd t}=-J_y,
 \end{gathered}
\]
and $\ms T$ differs from the identity only in the rows
\[
 (\ms T\ms y)_{w_s}=w_s-\mathrm ik\delta^2V\kappa_sE_x,\qquad
 (\ms T\ms y)_{E_x}=E_x+\frac{\mathrm ik\delta^2V}{\epsilon_0}J_y.
\]
The cold-fluid equations conserve
\[
 \mathcal H_0(\ms y)=\sum_s\rho_s\bigl(|u_s|^2+|w_s|^2\bigr)+\epsilon_0|E_x|^2+\epsilon_0|E_y-Vb|^2
 +\Bigl(\frac1{\mu_0}-\epsilon_0V^2\Bigr)|b|^2=\ms y^{\rm H}\ms H_0\ms y,
\]
that is, $\operatorname{Re}(\ms y^{\rm H}\ms H_0\ms L\ms y)=0$ for every $\ms y$. The entries that $\ms T$ adds to $\ms H_0$, $-\mathrm ik\delta^2V\kappa_s\rho_s$ in the rows $w_s$ and $\mathrm ik\delta^2V\kappa_s\rho_s$ in the row
$E_x$, are complex conjugates of each other, so that $\ms H_0\ms T$ is Hermitian. The form
$\mathcal H_{\Delta t}(\ms y)=\ms y^{\rm H}\ms H_0\ms T\ms y$ is therefore real, and the step conserves it,
\begin{align*}
 \mathcal H_{\Delta t}(\ms y^{n+1})-\mathcal H_{\Delta t}(\ms y^n)
 &=\operatorname{Re}\bigl[(\ms y^{n+1}+\ms y^n)^{\rm H}\ms H_0\ms T(\ms y^{n+1}-\ms y^n)\bigr]\\
 &=\frac{\Delta t}{2}\operatorname{Re}\bigl[(\ms y^{n+1}+\ms y^n)^{\rm H}\ms H_0\ms L(\ms y^{n+1}+\ms y^n)\bigr]=0.
\end{align*}
Completing the square gives
\[
 \begin{aligned}
 \mathcal H_{\Delta t}(\ms y)={}&\sum_s\rho_s|u_s|^2+\sum_s\rho_s\bigl|w_s-\mathrm ik\delta^2V\kappa_sE_x\bigr|^2
 +\epsilon_0\bigl[1-(kV\omega_p\delta^2)^2\bigr]|E_x|^2\\
 &+\epsilon_0|E_y-Vb|^2+\Bigl(\frac1{\mu_0}-\epsilon_0V^2\Bigr)|b|^2,
 \end{aligned}
\]
with $\omega_p^2=\epsilon_0^{-1}\sum_s\kappa_s^2\rho_s$. For $|V|<c$, $\mathcal H_{\Delta t}$ is therefore positive definite if and only if
$|kV|\omega_p\Delta t^2<4$, and the perturbations then remain bounded. For all $|k|\le k_*$, the largest modified wavenumber of the discrete curl, the condition holds if and only if
$k_*|V|\omega_p\Delta t^2<4$, that is, $\beta_i<4$. In the computations of Fig.~\ref{fig:tests-drift}, where $\omega_{pe}\Delta t\approx2.5$ and $1.25$, $\beta_i\approx0.76$ and
$0.19$, well below this bound.

The entry of $\ms T$ in the row $E_x$ comes from the drift of the charges and does not depend on the rotation field,
whereas its counterpart in the rows $w_s$ comes from \eqref{eq:frozen-b}. With $b^{\rm rot}=b^n$ in place of
\eqref{eq:frozen-b}, this counterpart is missing, and the row $w_s$ of the step instead reads
$w^{n+1}_s-w^n_s=\Delta t\,(\ms L\bar{\ms y})_{w_s}+2\mathrm ik\delta^2V\kappa_s\bar E_x$. The matrix $\ms H_0\ms T$ is then not
Hermitian, and the step conserves neither $\mathcal H_0$ nor $\mathcal H_{\Delta t}$. At $\Delta t=0.025$, the drift supplies energy to the perturbations, which grow and heat the electrons (Fig.~\ref{fig:tests-drift}(a)).

\section{Proof of Proposition~\ref{prop:gauss}}\label{app:proofs}

\begin{proof}
(i) For a charged species $s$ and a cell $\Omega_i$ that it occupies, let $\zeta_{s,i}$ be the vector of the coefficients
$\zeta_{s,i,\ell}$ at the nodes of $\Omega_i$. The second term of \eqref{eq:gauss-cost} is then
$\sum_{s,i}\frac{k_BT_{s,i}}{2m_s}\zeta_{s,i}^\tr\Gamma_{s,i}\zeta_{s,i}$ with
\[
 (\Gamma_{s,i})_{\ell\ell'}=\sum_{k\in\mathcal P_{s,i}}m_k\lambda_\ell(\vb x_k)\lambda_{\ell'}(\vb x_k)
 +\int_{\Omega_i}\rho^{h,c}_s\lambda_\ell\lambda_{\ell'}\dd\vb x,
\]
the Gram matrix of the nodal functions $\lambda_\ell$ with respect to the masses of the particles and the wave in
$\Omega_i$, which is symmetric and positive semidefinite.

The problem has a feasible point. As $\ms r_\parallel$ lies in the range of $\ms D$, there is a field $\ms e_F$ with
$\epsilon_0\ms D\ms e_F=\epsilon_0\ms D\ms e^c-\ms r_\parallel$. With $\zeta=0$, the masses are unchanged. Then
$\ms Q^{n+1}=\ms Q$, $\W^{n+1}_{s,i}=\W^c_{s,i}$ and $\epsilon_0\ms D\ms e_F-\ms Q^{n+1}=\ms r-\ms r_\parallel=\ms r_\perp$, so that
$(\ms e_F,0)$ satisfies \eqref{eq:gauss-multiplier} and \eqref{eq:gauss-constraints}.

The cost depends on $\zeta$ only through $\eta_{s,i}=\Gamma_{s,i}^{1/2}\zeta_{s,i}$, and in the variables $(\ms e^{n+1},\eta)$
it reads
\[
 \mathcal J=\frac{\epsilon_0}{2}(\ms e^{n+1}-\ms e^c)^\tr\ms M_E(\ms e^{n+1}-\ms e^c)
 +\sum_{s,i}\frac{k_BT_{s,i}}{2m_s}|\eta_{s,i}|^2,
\]
which is strictly convex and coercive, as $\ms M_E$ is positive definite and $T_{s,i}>0$. The feasible set is a nonempty
polyhedron, and so is its image under the linear map $(\ms e^{n+1},\zeta)\mapsto(\ms e^{n+1},\eta)$. On this closed
convex set, $\mathcal J$ has exactly one minimizer, and every feasible point that the map sends to it minimizes
\eqref{eq:gauss-cost}. The minimization therefore has a solution. Two solutions have the same $\ms e^{n+1}$ and the same
$\eta$, so that their vectors $\zeta_{s,i}$ differ by vectors $y$ with $\Gamma_{s,i}y=0$.

This difference does not change the corrected state. The state \eqref{eq:gauss-update} and the charges $\ms Q^{n+1}$ depend
on $\zeta_{s,i}$ only through the values of $\zeta_s$ at the particles of $\mathcal P_{s,i}$ and through the product
$\zeta_sh^c_s$ on $\Omega_i$. For $z=\sum_\ell y_\ell\lambda_\ell$,
\[
 0=y^\tr\Gamma_{s,i}y=\sum_{k\in\mathcal P_{s,i}}m_k\,z(\vb x_k)^2+\int_{\Omega_i}\rho^{h,c}_s\,z^2\dd\vb x,
\]
so that $z$ vanishes at the particles of $\mathcal P_{s,i}$ and wherever $\rho^{h,c}_s>0$. As $h^c_s=\M[\W^{h,c}_s]$ is
proportional to $\rho^{h,c}_s$, the product $zh^c_s$ vanishes on $\Omega_i$, and all solutions give the same corrected
state.

(ii) Every solution satisfies \eqref{eq:gauss-constraints}. Its first equation states that the Gauss defect after the
correction is $\ms r_\perp$, and its second that the total mass, momentum and energy of every charged species are those
before the correction.

(iii) The total energy is the sum of the energies of the species and the electric and magnetic energies. The correction
leaves $\ms b$ unchanged, and by (ii) it conserves the energy of every charged species. It therefore changes the total
energy by the change of the electric energy.
\end{proof}

\bibliographystyle{elsarticle-num}
\bibliography{refs}
\end{document}